\documentclass[useAMS,usenatbib]{mn2e}
\usepackage[draft]{graphicx}

\title[Very Long-term Optical Variability of HMXBs in the SMC]
       {Very Long-term Optical Variability of High Mass X-ray Binaries in the SMC}

\author[A. F. Rajoelimanana, P. A. Charles, \& A. Udalski]{A. F. Rajoelimanana$^{1,2}$\thanks{E-mail: andry@saao.ac.za}, P. A. Charles$^{1,3}$\thanks{E-mail: pac@saao.ac.za}, and A. Udalski$^{4}$\\
$^{1}$South African Astronomical Observatory, P.O. Box 9, Observatory, 7935, South Africa \\
$^{2}$University of Cape Town, Private Bag X3, Rondebosch, 7701, South Africa \\
$^{3}$School of Physics and Astronomy, Southampton University, Southampton SO17 1BJ \\
$^{4}$Warsaw University Observatory, Aleje Ujazdowskie 4, 00-478 Warsaw, Poland}

\begin{document}

\date{20 December 2010, accepted for publication in MNRAS}

\pagerange{\pageref{firstpage}--\pageref{lastpage}}\pubyear{2010}

\maketitle

\label{firstpage}

\begin{abstract}
We have studied the very long-term temporal properties of the optical emission from Be X-ray binaries (BeX) in the Small Magellanic Cloud over a $\sim$ 16 yr baseline, using light curves from the MACHO and OGLE databases. All the BeX in our sample display superorbital variations, many of them quasi-periodic on timescales of $\sim$ 200-3000 d. These long-term variations are believed to be related to the formation and depletion of the circumstellar disc around the Be star and we compare and contrast their behaviour with that of the LMC's prototypical BeX, A0538-66. The great majority of sources show a correlation of outburst amplitude with brightness (the opposite to that seen in A0538-66) although the amplitudes are mostly small ($\le$ 0.1 mag). We suggest this is an orbital inclination effect. In addition, we have also detected many of their optical orbital periodicities, visible as a series of precisely regular outbursts. Furthermore, the amplitude of these periodic outbursts can vary through the long-term superorbital cycle, and we discuss mechanisms which can produce this effect, as well as examining an apparent correlation between these periodicities. As a by-product of this variation survey we have compiled a list of all the reported SMC BeX orbital and superorbital periodicities at optical and X-ray wavelengths.

\end{abstract}


\section{Introduction}
\label{intro}

In high-mass X-ray binaries (HMXBs), a compact object (usually a neutron star) accretes mass from a massive early type O-B  star. Conventionally, they are subdivided into two groups, the supergiant X-ray binaries (SgXRB) and Be/X-ray binaries (BeX). These massive X-ray binaries are particularly numerous in the Small Magellanic Cloud (SMC) \citep[e.g.][]{coe05}.

From extrapolation of the Milky Way's population of 65 HMXBS and based on the Milky Way/SMC mass ratio being $\approx$ 50, one would expect to find only one or two HMXBs in the SMC. Remarkably, 59 of these systems have now been detected, with another surprising result being that all of them are BeX systems, with only one exception, the supergiant system SMC X-1 \citep{coe05,mcgowan07}.

A BeX consists of a neutron star orbiting a Be star in a wide (period $\sim$ months) and eccentric orbit. For reasons not fully understood, Be stars rotate very rapidly, leading to outflows from the Be star that form an equatorial disc around it. If sufficiently extended, the neutron star can penetrate this equatorial disc during each periastron passage, giving rise to periodic outbursts that can be observed over a wide range of wavelengths (optical, X-ray,...) \citep[e.g.][]{okneg01}. A review of basic properties of BeX can be found in \citet{neg98} and \citet{coe05}.

Several authors have investigated the extensive optical light curves of these sources available from the MACHO and OGLE projects in search of evidence for orbital modulations \citep{alcock96,udalski97}. In addition to the orbital variations usually seen as a series of periodic outbursts, these sources can show long-term superorbital modulations with timescales of hundreds of days to years. The prototype for this behaviour is the highly luminous ($\ge$ 10$^{39}$ ergs$^{-1}$ at its peak) LMC BeX, A0538-66, which has a remarkably stable superorbital modulation of 420.82 days, compared to its well-established orbital period of 16.65 days \citep[hereafter MC03]{alcock01,mcgowan03}. It was suggested by MC03 that these long-term modulations were a result of the formation and depletion of the circumstellar disc around the donor star which gives rise to the Be phenomenon. Given the known long-term variations in the Be systems, this discovery raised the question of whether such behaviour was ubiquitous, or somehow related to the presence of the X-ray source.

Consequently, in this paper, we investigate this superorbital behaviour in much greater detail within the now substantial BeX population by exploiting the MACHO and OGLE project databases. Together these provide a 16 year baseline for LMC/SMC studies of X-ray binary counterparts. Table~\ref{tab:listBe} gives the position (RA and Dec) as well as alternative name of all known SMC X-ray pulsars. We use the naming convention of \citet{coe05}, where SXP`x' is the SMC X-ray Pulsar with a `x' s pulse period. Table~\ref{tab:listcounterpart} details the MACHO and OGLE counterparts of these SMC X-ray pulsars, their OGLE positions (RA and Dec), all previously reported X-ray and optical orbital periods, and their X-ray luminosities.

\section{Optical light-curves from the MACHO and OGLE projects}
\label{long}

The MACHO (MAssive Compact Halo Objects) observations were made with the 1.27 m telescope located at Mount Stromlo Observatory, Australia \citep{alcock96}. It provides photometry in two passbands, a red band which we refer to as $\mathcal{R}$-band ($\sim$ 6300-7600 \AA , a slightly longer effective wavelength than Johnson R) and a blue band we refer to as $\mathcal{V}$-band ($\sim$ 4500-6300 \AA, slightly shorter effective wavelength than Johnson V). The data were taken during the interval 1993 to 2000 and are available at the MACHO website\footnote{wwwmacho.anu.edu.au}. The data are in instrumental magnitudes.

The OGLE (Optical Gravitational Lensing Experiment) projects have used the 1.3 m  Warsaw telescope at Las Campanas Observatory, Chile. Phase II of the OGLE project \textbf{ran} from January 1997 to Dec 2000 (\citet{udalski97}; \citet{zebrun01}), and phase III from 2001 until 3 May 2009. The photometric data for the OGLE projects are taken in the I-band. OGLE II data can be downloaded from the OGLE II website\footnote{ogle.astrouw.edu.pl}. OGLE III data for optical counterparts of X-ray sources located in the fields observed regularly by the OGLE-III survey are available at the X-ray variables OGLE monitoring (XROM) website\footnote{ogle.astrouw.edu.pl/ogle3/xrom/xrom.html} \citep{udalski08}.

\subsection{Calibrations}

We used the merged OGLE II and OGLE III photometry which are relative to the final OGLE III photometry. The details of the OGLE-III calibrations which based on OGLE II photometry can be found in \citet{udalski08a}. For the MACHO data we determined what offset is needed to align them with the OGLE II data as both projects observed the SMC from 1997 to 2000, this provides substantial overlap for demonstrating their compatibility.

\subsection{Temporal analysis}

The combined lightcurves($\sim$16 years) were then analysed using the Starlink PERIOD\footnote{www.starlink.rl.ac.uk/star/docs/sun167.htx/sun167.html} time-series analysis package designed to search for periodicities in unevenly sampled data. It has a variety of periodicity search options, including the Lomb-Scargle periodogram \citep[LS periodogram,][]{lomb76,scargle82}, phase dispersion minimization \citep[PDM,][]{stelling78}, and CLEAN algorithm \citep{roberts87}. The PERIOD package also has the capability to detrend and to fold any data with a given period.

\subsection{Secular variability in combined datasets}

In performing the temporal analyses reported here, it became clear that there were apparently significant modulations present in a few cases, but these were of low amplitude ($\sim$1\% peak-to-peak).  Given that our data were generated from independent telescopes with different detector/filter combinations, we decided to study randomly selected field stars in order to ascertain the properties of any secular, long-term variability that this might induce.  For those of our SMC fields containing BeX targets that exhibited low-amplitude modulations, we selected $\sim$ 25 random, presumed ``constant'' stars for analysis in the same way as our targets.

\begin{figure}
\scalebox{1}{\includegraphics{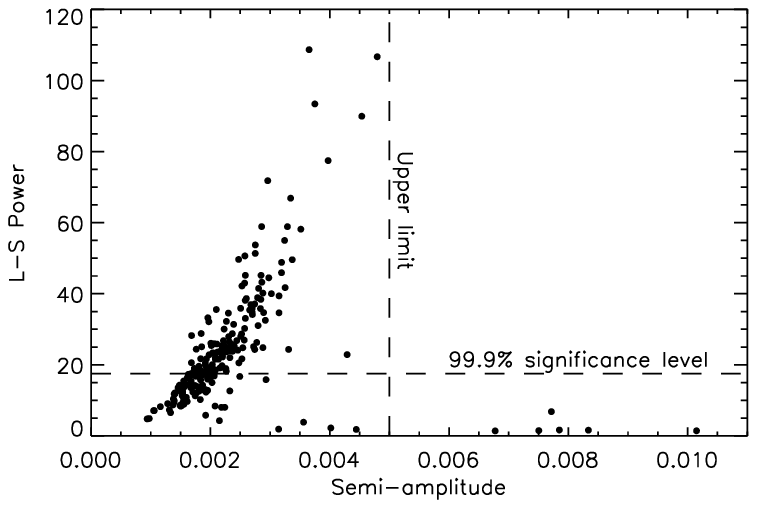}}
\caption{The peak L-S power observed for 221 random stars selected from fields where low amplitude modulations had been detected in our BeX target. These are plotted against the associated semi-amplitude and allow us to set a limit of 5 mmag for any real modulation. The 99.9\% confidence level for significant periods in terms of the L-S power are based on white noise calculations.}
\label{plotpowerampl} 
\end{figure}

Figure~\ref{plotpowerampl} summarises our results by displaying the peak L-S power observed for each random star plotted against the associated semi-amplitude (obtained by folding the data on that peak period and fitting a sine wave).  Interestingly, approximately half of these random stars produced apparently significant peaks in the power spectrum (as defined by the white noise 99.9\% confidence level).  However, the amplitudes of these modulations are all low ($<$1\% peak-to-peak), and are all below the lowest amplitude modulations seen in any of our target BeX systems.

Nevertheless,  where there is no independent detection of periodicities in our BeX targets (in either X-ray or optical) we have elected to report apparent detections at these low amplitudes as upper limits in table~\ref{tab:listres}. This is only true in 6 cases, and has no impact on any of our subsequent results or conclusions.

It is possible that some of the highest power peaks in figure~\ref{plotpowerampl} are real, long-term modulations in these field early-type stars in the SMC. However, their low amplitudes will require more extensive observations and careful analysis for confirmation and follow-up. This could be an interesting research project in its own right.

\begin{table}
\centering
\caption{List of all known SMC X-ray pulsars.}
\vspace{0.5cm}
\scriptsize{
\begin{tabular}{lllr}\hline\hline
\\
\multicolumn{1}{c}{\(\bf {Short}\)}&
\multicolumn{1}{c}{\(\bf {RA}\)}&
\multicolumn{1}{c}{\(\bf {Dec}\)} &
\multicolumn{1}{c}{\(\bf {Alternative^{\dagger}}\)} \\

\multicolumn{1}{c}{\(\bf {ID}\)}&
\multicolumn{1}{c}{\(\bf {(2000)}\)}&
\multicolumn{1}{c}{\(\bf {(2000)}\)} &
\multicolumn{1}{c}{\(\bf {Name}\)} \\

\\
\hline
\\

SXP0.09	 & 	00:42:35.4	 & 	-73:40:34	 & AX J0043-737		 	  	\\ 
SXP0.72	 & 	01:17:05.5	 & 	-73:26:32	 & 2U 0115-737,~SMC X-1                 \\  
SXP0.92	 & 	00:45:35.0	 & 	-73:19:02	 & PSR J0045-7319	 	  	\\ 
SXP2.16	 & 	01:19:51.0	 & 	-73:11:48	 & XTE J0119-731          	  	\\ 
SXP2.37	 & 	00:54:36.2	 & 	-73:40:35	 & H 0053-739,~SMCX-2    	        \\ 
SXP2.76	 & 	00:59:11.3	 & 	-71:38:45	 & RX J0059.2-7138   	          	\\ 
SXP3.34	 & 	01:05:09.7	 & 	-72:11:46	 & AX J0105-722,~[MA93] 1506            \\ 
SXP4.78	 & 	00:52:06.6	 & 	-72:20:44	 & XTE J0052-723,~[MA93] 537     	\\ 
SXP6.85	 & 	01:03:24.0	 & 	-72:43:00	 & XTE J0103-728    		  	\\ 
SXP7.78	 & 	00:52:07.7	 & 	-72:25:50	 & 2S 0050-727,~[MA93] 531 		\\ 
SXP7.92	 & 	01:01:56.0	 & 	-72:32:36	 & AZV285		  	  	\\ 
SXP8.02  &	01 00 43.1	 & 	-72 11 33    	 & PSR J0100-7211			\\
SXP8.9	 & 	00:51:53.0	 & 	-72:31:45	 & RX J0051.8-7231,[MA93]506 		\\
SXP9.13	 & 	00:49:13.8	 & 	-73:11:36	 & RX J0049.2-7311		  	\\
SXP11.5  &	01 04 41.4	 &	-72 54 04    	 & IGR J01054-7253 			\\
SXP15.3	 & 	00:52:13.9	 & 	-73:19:13	 & RX J0052.1-7319,[MA93]552	   	\\ 
SXP16.6	 & 	00:50:00.0	 & 	-73:16:00	 & XTE J0050-732\#1    	  	  	\\
SXP18.3	 & 	00:49:11.4	 & 	-72:49:37    	 & XMMJU004911.4-724939	  	  	\\ 
SXP22.1	 & 	01:17:41.4	 & 	-73:30:49	 & RX J0117.6-7330   		  	\\ 
SXP25.5	 & 	00:48:14.1	 & 	-73:10:03	 & XMMU J004814.1-731003	  	\\ 
SXP31.0	 & 	01:11:08.4	 & 	-73:16:46	 & XTE J0111.2-7317                    	\\ 
SXP34.0	 & 	00:55:27.7	 & 	-72:10:59	 & CXOU J005527.9-721058 	  	\\ 
SXP46.4	 & 	00:51:00.0	 & 	-73:18:00	 & XTE SMC46             	  	\\ 
SXP46.6	 & 	00:53:55.0	 & 	-72:26:47	 & 1WGA J0053.8-7226     	  	\\ 
SXP51	 & 	00:50:00.0	 & 	-73:16:00	 & XTE J0050-732\#2       	  	\\ 
SXP59 	 & 	00:54:56.1	 & 	-72:26:27	 & XTE J0055-724,~[MA93] 810        	\\ 
SXP65.8  &	01:07:12.6	 &	-72:35:33	 & CXOU J010712.6-723533		\\
SXP74.7	 & 	00:49:02.5	 & 	-72:50:52	 & AX J0049-729          	  	\\ 
SXP82.4	 & 	00:52:09.1	 & 	-72:38:03	 & XTE J0052-725			\\ 
SXP89.0	 & 	00:53:54.0	 & 	-72:26:42	 & XTE POSITION A        	  	\\ 
SXP91.1	 & 	00:50:55.8	 & 	-72:13:55	 & RX J0051.3-7216,~[MA93] 413     	\\ 
SXP95	 & 	00:53:24.0	 & 	-72:49:18	 & XTE SMC95             	  	\\ 
SXP101	 & 	00:57:26.8	 & 	-73:25:02	 & AX J0057.4-7325       	  	\\ 
SXP138	 & 	00:53:23.8	 & 	-72:27:15	 & CXOUJ005323.8-722715			\\ 
SXP140	 & 	00:56:05.2	 & 	-72:22:00	 & XMMUJ005605.2-722200			\\ 
SXP144	 & 			 & 		     	 & XTE SMC144s           	  	\\ 
SXP152	 & 	00:57:50.3	 & 	-72:07:56	 & CXOUJ005750.3-720756   		\\ 
SXP164	 & 	01:05:00.0	 & 	-72:06:21	 & XTE POSITION B        	  	\\ 
SXP169	 & 	00:52:59.2	 & 	-71:57:58	 & RX J0052.9-7158,~[MA93] 623      	\\ 
SXP172	 & 	00:51:52.2	 & 	-73:10:33	 & RX J0051.9-7311,~[MA93] 504       	\\ 
SXP202A	 & 	00:59:21.0	 & 	-72:23:16	 & 1XMMU J005921.0-722317	  	\\ 
SXP202B	 & 	00:59:28.6	 & 	-72:37:03	 & XMMUJ005929.0-723703		 	\\ 
SXP264	 & 	00:47:23.4	 & 	-73:12:27	 & RX J0047.3-7312,~[MA93] 172      	\\ 
SXP280	 & 	00:57:48.4	 & 	-72:02:42	 & RX J0057.8-7202,~[MA93]1036      	\\ 
SXP293	 & 	00:58:11.7	 & 	-72:30:50	 & RX J0058.2-7231       	  	\\ 
SXP304	 & 	01:01:01.1	 & 	-72:06:57	 & RX J0101.0-7206,~[MA93]1240      	\\ 
SXP323	 & 	00:50:44.7	 & 	-73:16:05	 & RX J0050.7-7316,~[MA93]387 	  	\\ 
SXP327	 & 	00:52:52.2	 & 	-72:17:14	 & XMMU J005252.1-721714.7       	\\ 
SXP348	 & 	01:03:13.9	 & 	-72:09:14	 & AX J0103-722,~[MA93] 1367         	\\ 
SXP455	 & 	01:01:20.5	 & 	-72:11:18	 & RX J0101.3-7211,~[MA93]1257      	\\ 
SXP504   & 	00:54:55.8	 & 	-72:45:10	 & AX J0054.8-7244,~[MA93]809       	\\ 
SXP564	 & 	00:57:36.2	 & 	-72:19:34	 & CXOUJ005736.2-721934   	 	\\ 
SXP645	 & 	00:55:35.1	 & 	-72:29:06	 & XMMU J005535.2-722906		\\ 
SXP701	 & 	00:55:18.4	 & 	-72:38:51	 & XMMU J005517.9-723853 	  	\\ 
SXP726	 & 	01:05:55.3	 & 	-72:03:47	 & RX J0105.9-7203,~[MA93]1557      	\\ 
SXP755   & 	00:49:42.0	 & 	-73:23:15	 & RX J0049.7-7323,~[MA93]315       	\\ 
SXP893	 &	00:49:29.8	 &	-73:10:58    	 &					\\
SXP967	 &	01:02:06.7       &	-71:41:15     	 & CXOU J010206.6-714115		\\
SXP1323  & 	01:03:37.5	 & 	-72:01:33	 & RX J0103.6-7201,~[MA93]1393      	\\ 

\\
\hline
\multicolumn{4}{p{7.5cm}}{$^{\dagger}$ [MA93] is related to the catalogue of emission-line stars and PNe by \citet{meys93}.}

\label{tab:listBe}
\end{tabular}
}
\end{table}

\begin{table*}
\centering
\caption{MACHO and OGLE counterparts of SMC X-ray pulsars.}
\vspace{0.5cm}
\scriptsize{
\begin{tabular}{lllcllrlrrll}\hline\hline \\
\multicolumn{1}{c}{\(\bf {Short}\)}&
\multicolumn{1}{c}{\(\bf {V}\)} &
\multicolumn{1}{c}{\(\bf {Spec}\)}&
\multicolumn{1}{c}{\(\bf {Lum}\)}&
\multicolumn{1}{c}{\(\bf {MACHO}\)}&
\multicolumn{2}{c}{\(\bf {OGLE~II}\)}&
\multicolumn{2}{c}{\(\bf {OGLE~III}\)}&
\multicolumn{1}{c}{\(\bf {P_{{\mbox{X-ray}}}^{\dagger}}\)}&
\multicolumn{1}{c}{\(\bf {P_{{\mbox{opt}}}^{\ddagger \star}}\)}&
\multicolumn{1}{c}{\(\bf {Log~L_{\mbox{X}}(max)} \)}	\\

\multicolumn{1}{c}{\(\bf {ID}\)}&
\multicolumn{1}{c}{\(\bf {}\)}&
\multicolumn{1}{c}{\(\bf {type}\)} &
\multicolumn{1}{c}{\(\bf {class}\)}&
\multicolumn{1}{c}{\(\bf {catalogue}\)} &
\multicolumn{1}{c}{\(\bf {fields}\)}&
\multicolumn{1}{c}{\(\bf {ID}\)}&
\multicolumn{1}{c}{\(\bf {fields}\)}&
\multicolumn{1}{c}{\(\bf {ID}\)}&
\multicolumn{1}{c}{\(\bf {(d)~}\)}&
\multicolumn{1}{c}{\(\bf {(d)~}\)}&
\multicolumn{1}{c}{\(\bf {(erg.s^{-1})} \)}	\\
\\
\hline
\\
SXP0.09	&	16.9	&			&		&			&	SMC-SC3		&	6	&	SMC128.4	&	118	&		&	  					&	34.9	[0.7-10~kev]	\\
SXP0.92	&	16.1	&	B0.5-B2		&	IV-V	&	212.15676.64	&	SMC-SC3		&	197970	&	SMC125.3	&	19461	&		&	51{\tiny $^{{\mbox{[1][R]}}}$}		&				\\
SXP2.37	&	16.3	&	O9.5		&	III-V	&			&			&		&	SMC107.5	&	25	&		&						&	38.6	[2-25~kev]	\\
SXP2.76	&	14.0	&	B1-B1.5		&	II-III	&			&			&		&	SMC109.2	&	5	&		&	82.1{\tiny $^{{\mbox{[2][O]}}}$}	&	37.7	[0.1-2~kev]	\\
SXP3.34	&	15.6	&	B1-B2		&	III-V	&	206.16890.17	&	SMC-SC10	&	102553	&			&		&		&						&	35.1	[0.1-2~kev]	\\
SXP4.78	&	15.8	&	B0-B1		&	V	&	207.16146.9	&			&		&			&		&	34.1	&	23.9{\tiny $^{{\mbox{[3][M]}}}$}	&	37.8	[3-10~kev]	\\
SXP6.85	&	14.5	&	O9.5-B0		&	IV-V	&	206.16768.5	&	SMC-SC9		&	146936	&			&		&	112	&	114{\tiny $^{{\mbox{[4][OM]}}}$}	&	37.5	[2-30~kev]	\\
SXP7.78	&	14.9	&	B1-B1.5		&	IV-V	&	208.16088.4	&	       		&		&	SMC101.3	&	33277	&	44.9	&	44.8{\tiny $^{{\mbox{[5][OM]}}}$}	&	37.7	[2-10~kev] 	\\
SXP7.92	&	13.9	&	O9 		&		&	207.16714.4	&	SMC-SC9		&	114259	&	SMC110.5	&	5	&		&	36.8{\tiny $^{{\mbox{[14][O]}}}$}	&	36.8	[0.3-10~kev]	\\
SXP8.9	&	14.8	&	O9.5-B0		&	IV-V	&	208.16087.9	&	SMC-SC6		&	85614	&	SMC101.3	&	4737	&	28.4	&	33.4{\tiny $^{{\mbox{[6][OM]}}}$}	&	36.1	[0.1-2~kev]	\\
SXP9.13	&	16.5	&	B1-B3		&	IV-V	&	212.15906.2446	&	SMC-SC5		&	111490	&	SMC100.7	&	63223	&	77.2	&	40.1{\tiny $^{{\mbox{[8][O]}}}$}	&	35.6	[0.7-10~kev]	\\
SXP15.3	&	14.6	&	O9.5-B0		&	III-V	&	212.16075.13	&	SMC-SC6		&	99923	&	SMC100.1	&	48026	&	28	&	75.1{\tiny $^{{\mbox{[8][OM]}}}$}	&	37.1	[0.1-2~kev] 	\\
SXP18.3	&	15.6	&	B0-B2		&	V	&	208.15911.13	&	SMC-SC5		&	65500	&	SMC101.8	&	19552	&	17.7	&	17.7{\tiny $^{{\mbox{[15][O]}}}$}	&	37.5	[0.2-10~kev]	\\
SXP22.1	&	14.1	&	O9.5-B0		&	III-V	&			&			&		&	SMC116.1	&	3322	&		&						&	38.0	[0.1-2~kev]	\\
SXP25.5	&	15.2	&			&		&	212.15849.52	&	SMC-SC4		&	171264	&	SMC100.7	&	54315	&		&						&	35.3	[0.2-10~kev]	\\
SXP31.0	&	15.5	&	O9.5-B1		&	V	&			&			&		&	SMC116.6	&	33	&		&	90.4{\tiny $^{{\mbox{[2][O]}}}$}	&	38.3	[0.1-2~kev]	\\
SXP34.1	&	16.7	&	B2-B3		&	IV-V	&			&			&		&	SMC108.7	&	24657	&		&						&	35.0	[0.3-10~kev]	\\
SXP46.6	&	14.7	&	O9.5-B1		&	IV-V	&			&			&		&	SMC108.8	&	30	&	137	&	137{\tiny $^{{\mbox{[9][O]}}}$}		&	36.8	[0.1-2~kev]	\\
SXP59	&	15.2	&	O9		&	V	&	207.16259.23	&	SMC-SC7		&	70829	&	SMC105.5	&	35420	&	122	&	60.2{\tiny $^{{\mbox{[10][OM]}}}$}	&	37.6	[2-10~kev]	\\
SXP65	&	15.6	&	B1-B1.5		&	II-III	&			&			&     		&			&		&	110	&						&	35.8	[0.2-10~kev]	\\
SXP74.7	&	16.9	&	B3		&	V	&	208.15911.93	&	SMC-SC5		&	65517	&	SMC100.5	&	55158	&	61.6	&	33.4{\tiny $^{{\mbox{[11][O]}}}$}	&	36.7	[0.7-10~kev]	\\
SXP82.4	&	15.0	&	B1-B3		&	III-V	&	208.16085.24	&	SMC-SC6		&	77228	&	SMC101.2	&	12997	&	362	&						&	36.5	[0.3-10~kev]	\\
SXP91.1	&	15.0	&	B0.5		&	III-V	&	208.16034.5	&			&		&	SMC102.1	&	32	&	117	&	88.2{\tiny $^{{\mbox{[7][M]}}}$}	&	37.4	[0.1-2~kev]	\\
SXP101	&	15.6	&	B3-B5		&	Ib-II	&	211.16415.11	&			&		&	SMC106.7	&	15343	&	25.2	&	21.9{\tiny $^{{\mbox{[12][OM]}}}$}	&	36.0	[0.7-10~kev]	\\
SXP138	&	16.1	&	B1-B2		&	IV-V	&	207.16202.50	&			&		&	SMC101.3	&	38781	&	103	&	125{\tiny $^{{\mbox{[8][M]}}}$}		&	35.0	[0.3-10~kev]	\\
SXP140	&	15.8	&	B1		&	V	&	207.16374.21	&			&		&	SMC108.8	&	35838	&		&	197{\tiny $^{{\mbox{[6][M]}}}$}		&	34.6	[0.1-2~kev]	\\
SXP152	&	15.6	&	B1-B2.5		&	III-V	&			&			&		&	SMC108.3	&	36	&		&						&	35.6	[0.1-2~kev]	\\
SXP169	&	15.5	&	B0-B1		&	III-V	&			&			&		&	SMC102.2	&	13410	&	68.5	&	67.6{\tiny $^{{\mbox{[2][O]}}}$}	&	37.3	[0.1-2~kev]	\\
SXP172	&	14.4	&	O9.5-B0		&	V	&	212.16077.13	&	SMC-SC6		&	22749	&	SMC100.2	&	44100	&	70	&	69.9{\tiny $^{{\mbox{[6][O]}}}$}	&	36.0	[0.2-10~kev]	\\
SXP202A	&	14.8	&	B0-B1		&	V	&	207.16545.12	&	SMC-SC8		&	151891	&	SMC108.1	&	4929	&	91	&						&	35.5	[0.2-10~kev]	\\
SXP202B	&	15.6	&	B0-5		&	III	&	207.16541.15	&	SMC-SC8		&	139407	&	SMC105.3	&	29894	&		&						&	36.6	[0.2-10~kev]	\\
SXP264	&	15.8	&	B1-B1.5		&	V	&	212.15792.77	&	SMC-SC4		&	116979	&	SMC100.7	&	45007	&		&	49.1{\tiny $^{{\mbox{[10][O]}}}$}	&	36.2	[0.2-10~kev]	\\
SXP280	&	15.6	&	B0-B2		&	III-V	&			&			&		&	SMC108.3	&	16091	&	64.8	&	127{\tiny $^{{\mbox{[2][O]}}}$}		&	36.2	[0.1-2~kev]	\\
SXP293	&	14.9	&	B2-B3		&	V	&	207.16486.9	&	SMC-SC8		&	94537	&	SMC105.4	&	22335	&	151	&	59.7{\tiny $^{{\mbox{[7][OM]}}}$}	&	35.6	[0.2-10~kev]	\\
SXP304	&	15.7	&	B0-B2		&	III-V	&	206.16663.16	&	SMC-SC9		&	47428	&	SMC108.3	&	12190	&		&	520{\tiny $^{{\mbox{[6][M]}}}$}		&	36.1	[0.1-2~kev]	\\
SXP323	&	15.4	&	B0-B0.5		&	V	&	212.16019.30	&	SMC-SC5		&	180026	&	SMC100.2	&	48	&	116	&						&	35.4	[0.2-10~kev]	\\
SXP327	&	16.3	&			&		&	207.16147.60	&			&		&	SMC101.4	&	25097	&		&	45.9{\tiny $^{{\mbox{[16][O]}}}$}	&	35.5	[0.2-10~kev]	\\
SXP348	&	14.7	&	B0.5		&	IV-V	&	206.16776.17	&	SMC-SC9		&	173121	&	SMC108.8	&	33	&		&	93.9{\tiny $^{{\mbox{[6][O]}}}$}	&	36.0	[0.1-2~kev]	\\
SXP455	&	15.4	&	B0.5-B2		&	IV-V	&	206.16662.14	&	SMC-SC9		&	89374	&	SMC108.2	&	34801	&		&	74.7{\tiny $^{{\mbox{[7][OM]}}}$}	&	35.8	[0.7-10~kev]	\\
SXP504	&	14.9	&	B1		&	III-V	&	207.16254.16	&	SMC-SC7		&	47103	&	SMC105.7	&	36877	&	265	&	273{\tiny $^{{\mbox{[10][O]}}}$}	&	35.7	[0.3-10~kev]	\\
SXP564	&	15.9	&	B0-B2		&	IV-V	&	207.16432.1575	&	SMC-SC8		&	49531	&	SMC108.1	&	19293	&	151	&	95.3{\tiny $^{{\mbox{[7][M]}}}$}	&	34.8	[0.3-10~kev]	\\
SXP645	&	14.6	&	B0-B0.5		&	III-V	&	207.16315.28	&	SMC-SC7		&	137527	&	SMC105.5	&	35415	&		&						&	35.6	[0.2-10~kev]	\\
SXP701	&	15.8	&	O9.5		&	V	&	207.16313.35	&	SMC-SC7		&	129062	&	SMC105.6	&	36015	&		&	412{\tiny $^{{\mbox{[10][M]}}}$}	&	35.6	[0.2-10~kev]	\\
SXP726	&	15.6	&	B0.5-B3		&		&			&	SMC-SC10	&	137851	&	SMC113.3	&	10946	&		&						&	34.8	[0.1-2~kev]	\\
SXP755	&	14.9	&	O9.5-B0		&	III-V	&	212.15960.12	&	SMC-SC5		&	90506	&	SMC100.8	&	22903	&	389	&	394{\tiny $^{{\mbox{[7][M]}}}$}		&	35.9	[0.2-10~kev]	\\
SXP893	&	16.3	&	                &		&			&	SMC-SC5		&	111500	&	SMC100.7 	&	63151	&		&						&	34.5	[2-10~kev]	\\
SXP967	&	14.6	&	B0-0.5		&	III-V	&			&			&		&	SMC114.7	& 	39 	&		&						&	35.9	[0.2-10~kev]	\\
SXP1323	&	14.6	&	B0		&	III-V	&			&	SMC-SC10	&	37138	&	SMC113.6	&	27699	&		&	26.1{\tiny $^{{\mbox{[2][O]}}}$}	&	36.8	[0.2-10~kev]	\\
\hline
\multicolumn{12}{p{17cm}}{[1]: \citet{kaspi93} ; [2]: \citet{schmidtke06}; [3]: \citet{coe05} ; [4]: \citet{mcgowan08} ; [5]: \citet{cowley04} ; [6]: \citet{schmidtkecow06} ; [7]: \citet{schmidtke04} ; [8]: \citet{edge05a} ; [9]: \citet{mcgowan08} ; [10]: \citet{schmidtke05b} ; [11]: \citet{schmidtke07b} ; [12]: \citet{mcgowan07} ; [13]: \citet{edge05b}; [14]: \citet{coe09}; [15]: \citet{schurch09}; [16]: \citet{coe08}}\\
\\
\multicolumn{12}{p{17cm}}{$^{\dagger}$X-ray orbital period from \citet{galache08} paper.}\\
\multicolumn{12}{p{17cm}}{$^{\ddagger}$Previously reported optical orbital period.}\\
\multicolumn{12}{p{17cm}}{$^{\star}$Source of P$_{opt}$: M for MACHO project, O for OGLE, OM for both of the data, and R for radio.}
\label{tab:listcounterpart}
\end{tabular}
}
\end{table*}

\section{Temporal properties of individual sources}
\label{individual}

\subsection{SXP0.09 (AX J0043-737)}
The 87.58 ms X-ray pulsation from this source was detected by \citet{yoko00} during ASCA observations of the SMC on 1999 May 10-11. Optically identified as SMC-SC3 6 \citep{udalski08}, this source is not however present in the catalogue of emission-line stars and PNe \citep[hereafter MA93]{meys93}.

\begin{figure}
\scalebox{1}{\includegraphics{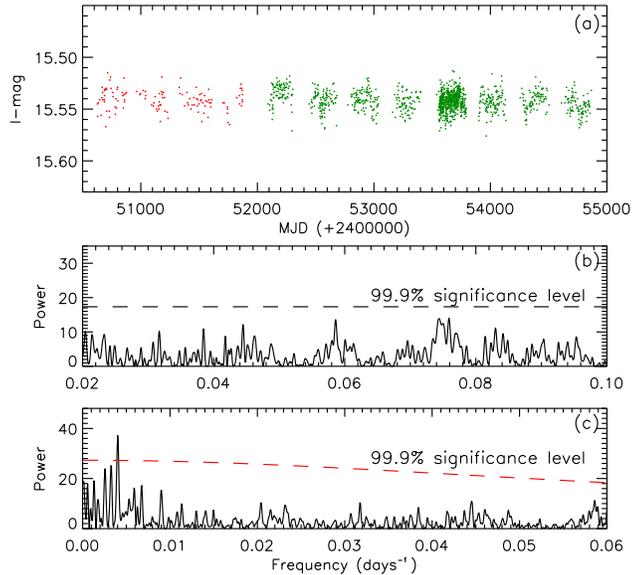}}
\caption{SXP0.09. (a): OGLE II (red) and OGLE III (black) light curves. (b) and (c): Lomb-Scargle power spectrum of the OGLE data with a significant peak at P=247 d, the red dashed line represents the 99.9\% confidence level relative to the ``red'' noise.}
\label{sxp0.09lc} 
\end{figure}

Figure~\ref{sxp0.09lc}a shows the combined OGLE II and OGLE III lightcurve of SXP0.09 (MACHO data is not available) which is very steady, to within $\pm$0.05 mag. Its orbital period has not yet been reported, but both the LS periodogram and PDM revealed a strong peak well above the 99.9 $\%$ confidence level relative to both white and red noise \footnote{Red noise is calculated using REDFIT \citep{schulz02} (www.ncdc.noaa.gov/paleo/softlib/redfit/redfit.html)} at a period of 247 $\pm$ 5 d ( Figure~\ref{sxp0.09lc}c). We did not find any significant shorter period in the periodogram (Figure~\ref{sxp0.09lc}b).

\subsection{SXP0.92 (PSR J0045-7319)}

SXP0.92 was first discovered by \citet{ables87} as a 0.926499 s radio pulsar. \citet{bell94} identified its optical companion as a 16th magnitude, 11 $M_{\odot}$, B1 main-sequence star. The orbital period (51 d) of PSR J0045-7319 was found by \citet{kaspi93} through the Doppler modulation of the pulse period.

\begin{figure}
\scalebox{1}{\includegraphics{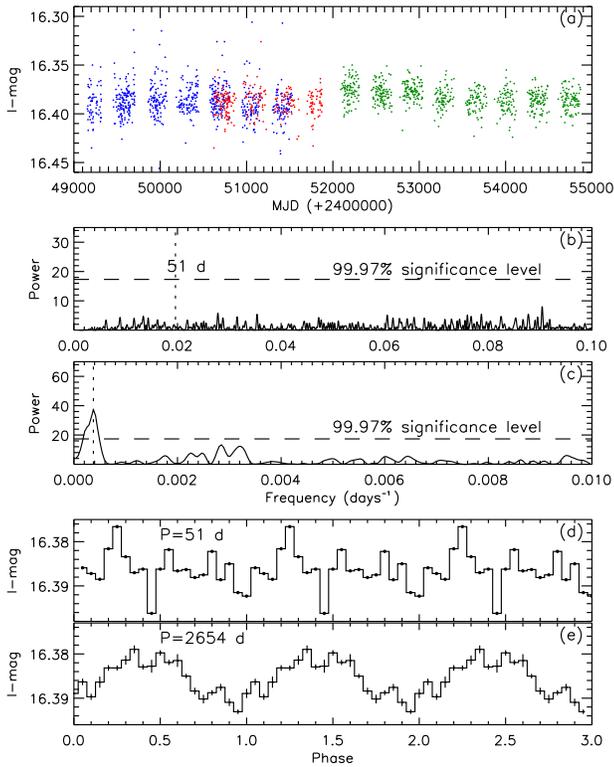}}
\caption{SXP0.92. (a): MACHO $\mathcal{R}$-band (blue), OGLE II (red) and OGLE III (green) light curves. (b) and (c): power spectrum of the combined data showing the peak at P=2654 d. (d) and (e): Combined data light curve folded on the 51 d presumed orbital period and the superorbital period of 2654 days.}
\label{sxp0.92lc} 
\end{figure}

Our combined light curve of SXP0.92 (Figure~\ref{sxp0.92lc}a) reveals no optical signature of the 51 d periodicity, but it does clearly display an almost sinusoidal $\sim$ 2600 d superorbital modulation. This appears as a strong peak in the LS periodogram at a frequency corresponding to 2654 $\pm$ 298 d (Figure~\ref{sxp0.92lc}c). The folded light curve on this period is shown in Figure~\ref{sxp0.92lc}d. Whilst this periodicity is very close to half the total observing time, the modulation is in fact clearly visible in the raw light curve (Figure~\ref{sxp0.92lc}a).

We did not find any significant peak at the previously reported 51 d period, but the data folded on this period shows a low amplitude ($\le$ 0.01 mag) modulation (Figure~\ref{sxp0.92lc}d).

\subsection{SXP2.37 (SMC X-2)}

SMC X-2 is a long-established X-ray source detected in the SMC by SAS-3 \citep{li77a}. The optical counterpart of SMC X-2 was identified by \citet{Pesch77} as a 14th magnitude OB star, and this identification was supported by spectroscopic observations \citep{allen77}.

\begin{figure}
\scalebox{0.46}{\includegraphics{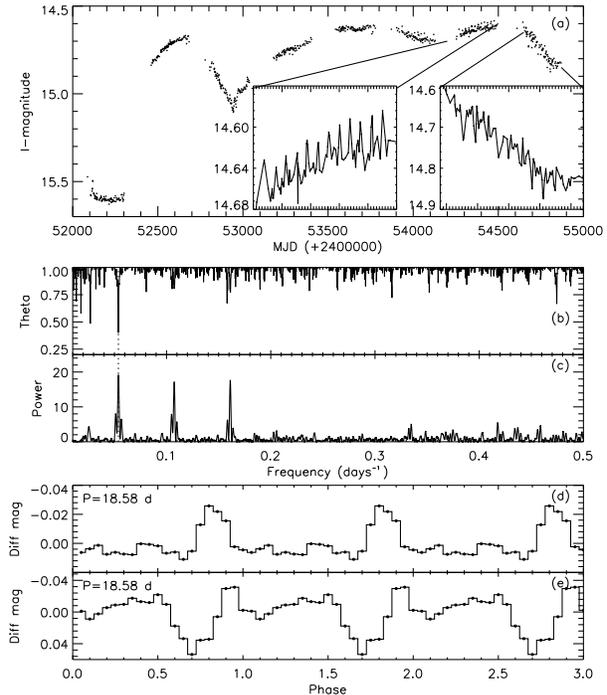}}
\caption{SXP2.37 : (a): OGLE III light curve with (inset) the blow-up of a section of data that displays a low amplitude modulation (outbursts and dips) at 18.58 d. (b) and (c): Periodogram (PDM and LS) of the OGLE III data with a highest peak at P=18.58 d, and its harmonics at 9.3 d and 6.2 d. (d) and (e): Light curve showing outbursts (MJD 53800- 54500) and dips (MJD 54600- 55000), respectively, folded on the presumed 18.58 d orbital period.}
\label{sxp2.37lc} 
\end{figure}

\citet{schurch08} found a variation in the OGLEIII data of SXP2.37 with a period of 18.62 $\pm$ 0.02 d, and proposed that it is the orbital period. However, \citet{schmidtke09} have suggested that this modulation is not orbital, but is instead the beat period between two non-radial pulsations (P=0.8592 d and 0.9008 d) of the primary star. We note that our OGLE III data has only one data point per night (only 3 nights have two). The OGLE III light curve of SXP2.37 clearly shows the 18.6 d variation (Figure~\ref{sxp2.37lc}a). We have detrended the OGLE III data and found peaks at 18.58$\pm$0.003 d, and its harmonics at 9.3 and 6.2 d (see Figure~\ref{sxp2.37lc}c), which is similar to those reported by \citet{schurch08}. The light curve folded on the 18.58 d period shows interesting evolution from apparent maxima to dips in each cycle (Figure~\ref{sxp2.37lc}d and e). This will be discussed further later.

\subsection{SXP2.76 (RX J0059.2-7138)}

The 2.76 s pulsation from RX J0059.2-7138 was detected with ROSAT \citep{hughes94}. \citet{schmidtke06} found the orbital period of SXP2.76 which is clearly seen as repeated, low-amplitude outbursts every 82.1 days.

\begin{figure}
\scalebox{0.46}{\includegraphics{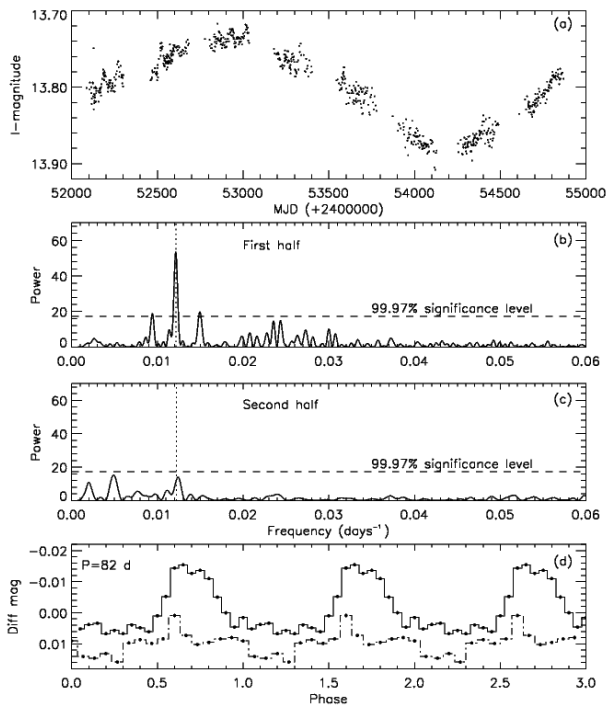}}
\caption{ SXP2.76: (a): OGLE III light curve. (b) and (c): power spectrum of the $1^{st}$ (before MDJ 53500) and $2^{nd}$ half of the detrended OGLE III data. (d): First (solid) and second (dashed) half of the OGLE III light curve folded on the period of 82 d.}
\label{sxp2.76lc}
\end{figure}

The source lies outside MACHO and OGLE II fields, but the OGLE III light curve for this X-ray pulsar (Figure~\ref{sxp2.76lc}a) shows an exceptionally clear sinusoidal variation of large ($\sim$0.2 mag) amplitude with a period of about 2800 d, which is comparable to the duration of the dataset. The first half of the detrended data (before MJD 53500) shows a very regular set of outbursts on the orbital period $\mathrm {P_{orb}}$ = 82$\pm$0.07 d (Figure~\ref{sxp2.76lc}b). In the second half, the amplitude of these outbursts is very low, and they completely disappear during global minimum. The power spectrum of the detrended second part of the OGLE III data does not show any significant peak (Figure~\ref{sxp2.76lc}c).

\subsection{SXP3.34 (AX J0105-722)}
\citet{coe05} identified the optical counterpart of AX J0105-722 as MA[93]1506 and found a period of 11.09 d in the MACHO data which they proposed to be the orbital period. However, \citet{schmidtke05a} again reported the presence of non-radial pulsations in the same MACHO data, the period being 1.099 d. They also suggested that the 11.09 d signal is an alias of this shorter period.

\begin{figure}
\scalebox{0.92}{\includegraphics{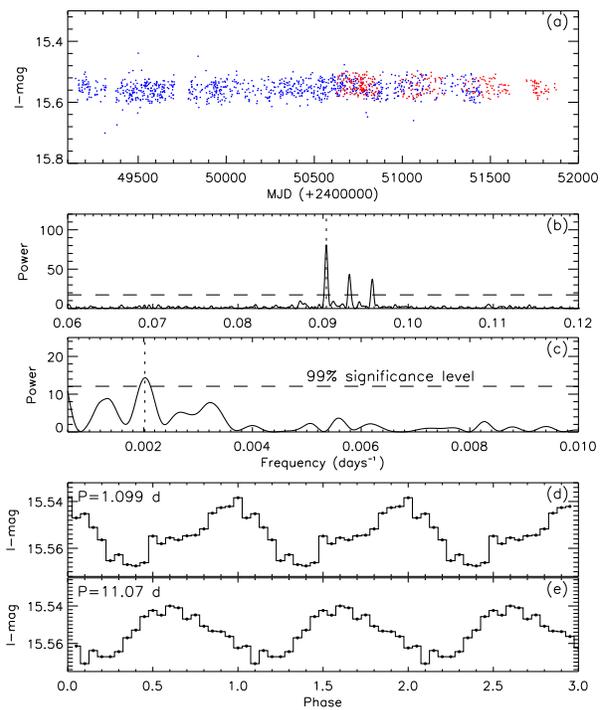}}
\caption{ SXP3.34: (a): MACHO $\mathcal{R}$-band (blue) and OGLE II (red) light curves. (b) and (c) : Power spectrum of the combined light curve showing peaks at 11.07 d and 495 d. (d) and (e): Light curve folded on the period of 1.099 d, 11.07 d respectively.}
\label{sxp3.34lc} 
\end{figure}

The combined MACHO and OGLE II light curve of SXP3.34 is presented in Figure~\ref{sxp3.34lc}a and is very stable to within $\pm$0.05 mag, although a several hundred day modulation is descernible. Our power spectrum shows a significant peak at 11.07$\pm$0.01 d (Figure~\ref{sxp3.34lc}b). The light curve folded on the 11.07 d period is shown in Figure~\ref{sxp3.34lc}e. The LS periodogram of the combined light curve shows a significant peak at 99\% confidence level with a much longer, but low amplitude super-orbital period of 495$\pm$2 d (Figure~\ref{sxp3.34lc}c).

\subsection{SXP6.85 (XTE J0103-728)}

X-ray pulsations at 6.85 s from XTE J0103-728 were first detected by \citet{corbet03a}. Subsequently, an accurate position was determined with XMM-Newton by \citet{haberl07} during an outburst in 2006 October, from which they identified the optical counterpart as a 14th magnitude Be star. \citet{schmidtke07a} have analyzed its lightcurve from MACHO observations and suggested an orbital period of 24.82 d. However, from the combined MACHO and OGLE datasets \citet{mcgowan07} found an optical period of 114 d which is similar to the X-ray period of 112.5 d reported by \citet{galache08} and is presumed to be orbital in origin.

\begin{figure}
\scalebox{0.46}{\includegraphics{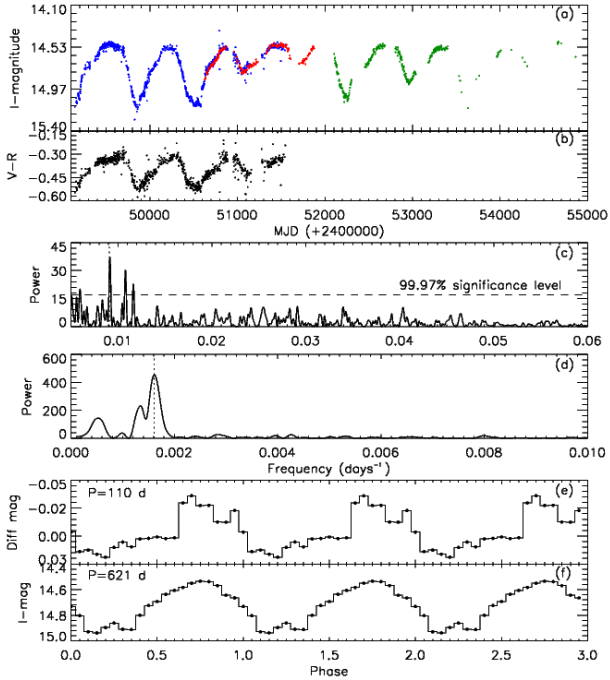}}
\caption{ SXP6.85: (a): MACHO and OGLE light curves. (b) MACHO colour variation ($\mathcal{V}-\mathcal{R}$). (c) and (d): Power spectrum of the detrended and combined light curves showing significant peaks at 110 d and 621 d respectively. (e) and (f): Light curve folded on the 110 d and 621 d.}
\label{sxp6.85lc} 
\end{figure}

We have detrended the complete datasets from MACHO, OGLE II and OGLE III, and found a significant peak at 110$\pm$0.2 d (Figure~\ref{sxp6.85lc}c), which is very similar to the previously reported periods. The folded light curve on this period shows a transient-like profile typical of BeX systems (Figure~\ref{sxp6.85lc}e).

The MACHO and OGLE light curves are presented in Figure~\ref{sxp6.85lc}a, and reveal a long-term variation of very large amplitude ($\geq 0.5~mag$). The power spectrum of the combined MACHO and OGLE light curves shows a large peak at 621 $\pm$4 d which is consistent with the `recurrence time' mentioned by \citet{schmidtke07a}.

The colour variation of SXP6.85 is also plotted, and the source is brighter when it is redder (Figure~\ref{sxp6.85lc}b). This is the opposite to that seen in A0538-66 where the source is bluer when it is brighter and strong outbursts occur only during optical minimum \citep{alcock01}. It is interesting to note that the color variation continues to increase until $\sim$ 100 days after the maximum brightness of the source, i.e., there is an offset (or lag) between the maximum of colour and brightness variations.

\citet{mcgowan07} have not found any correlation between X-ray and optical activity for this source and suggest that it could be a low eccentricity system.

\subsection{SXP7.78 (2S 0050-727)}

The optical counterpart of SXP7.78 is MA[93]531 or MACHO 208.16088.4 \citep{coe05}. From X-ray observations with RXTE, \citet{galache08} found a series of outbursts with a period of 44.92$\pm$0.06 d which is consistent with the presumed orbital period of 44.6 d found in the MACHO $\mathcal{R}$ lightcurve \citep{coe05}.

\begin{figure}
\scalebox{0.46}{\includegraphics{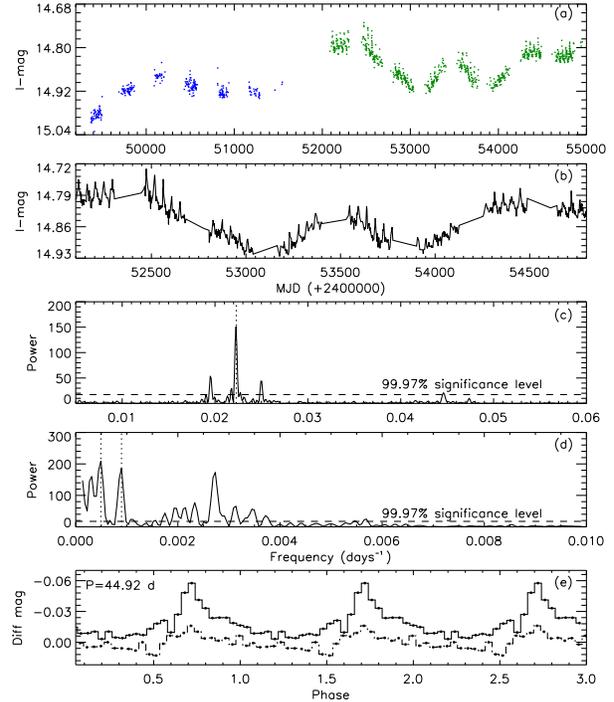}}
\caption{SXP7.78. (a): MACHO $\mathcal{R}$-band (blue) and OGLE III (green) light curves. (b): Blow-up of the OGLE III light curve. (c) and (d): Power spectrum of the detrended and combined data with maximum peaks at 44.92d, 1116 d and 2029 d (the peak at 0.002739 $d^{-1}$ is due to the annual sampling). (e): detrended light curve folded on the orbital period of 44.92 d when optically bright (solid line) and faint (dashed line).}
\label{sxp7.78lc} 
\end{figure}

The OGLE III lightcurve in Figure~\ref{sxp7.78lc}b shows clearly the regular orbital outbursts, as is also reflected in the power spectrum of the detrended data at a period of 44.92$\pm$0.2 d (Figure~\ref{sxp7.78lc}c). It is also clear that the amplitude of the outbursts vary with the source brightness.

The combined MACHO and OGLE III light curves of Figure~\ref{sxp7.78lc}a show substantial long-term variability, which is responsible for the LS power spectra peaks at 1116$\pm$56 d, and 2029$\pm$186 d. The peak at 0.002739 d$^{-1}$ (P=365 d) is due to the annual sampling of the data.

SXP7.78 is one of the BeX systems which shows clear optical periodic outbursts even when the source is at optical minimum in the superorbital cycle (see Figure~\ref{sxp7.78lc}b). Furthermore, the strength of its outbursts appears to vary with its optical brightness, being very strong at optical maximum, but always present throughout the 2029 d cyclic variation.

\subsection{SXP7.92 (AZV 285)}

The pulse period of 7.92 s of this source was discovered recently by \citet{corbet08} using RXTE/PCA. Its possible optical counterpart is discussed in \citet{coe09}, who identified it as AzV285 with an orbital period of 36.8 d. 

\begin{figure}
\scalebox{1}{\includegraphics{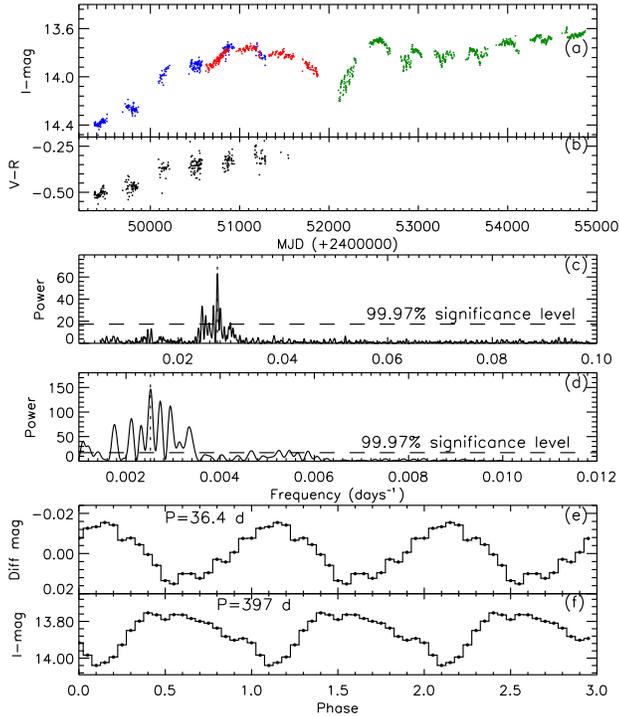}}
\caption{SXP7.92. (a): MACHO $\mathcal{R}$-band (blue), OGLE II (red) and OGLE III (green) light curves. (b): MACHO colour variation ($\mathcal{V}-\mathcal{R}$). (c) and (d): Power spectrum of the combined data showing peaks at P= 36.4 d and P= 397 d. (e) and (f): Folded light curves on the periods of 36.4 d and 397 d respectively.}
\label{sxp7.92lc} 
\end{figure}

The power spectrum of the long-term combined light curve reveals strong low-frequency power, with a peak at 397$\pm$2 d (Figure~\ref{sxp7.92lc}d) and the detrended light curves show a peak at 36.4$\pm$0.02 d (Figure~\ref{sxp7.92lc}c), very similar to the presumed orbital period. The folded light curves on these periods are presented in Figure~\ref{sxp7.92lc}e and \ref{sxp7.92lc}f, where the long-term one has a clear asymmetric profile.

\begin{figure}
\scalebox{0.46}{\includegraphics{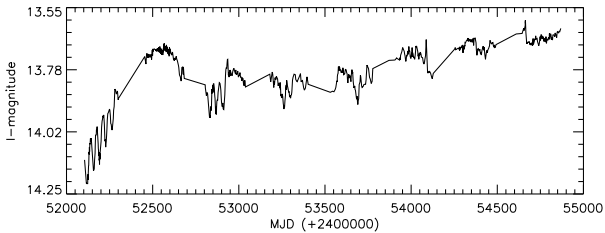}}
\caption{SXP7.92. Blow-up of the OGLE III light curve.}
\label{sxp7.92lc2} 
\end{figure}

The light curve of SXP7.92 exhibits large ($\sim$1 mag) variations and becomes redder when it is brighter (suggesting a low inclination system), and the orbital modulation is always present. But, as seen in the OGLE III light curve (Figure~\ref{sxp7.92lc2}), when the Be star is close to maximum (I$\sim$13.7), the strength of the normal periodic modulation becomes very weak. Optically at maximum, SXP7.92 is the brightest BeX system in the SMC.

\subsection{SXP8.88 (RX J0051.8-7231)}
 
This source was discovered to be an X-ray pulsar by \citet{israel97}. Its optical counterpart was proposed by \citet{haberlandpietsch04} to be MA[93]506. \citet{schmidtkecow06} found a weak period of 33.4 d in the early OGLE II data which they suggested as the orbital period. From RXTE observations of the SMC, \citet{corbet04a} suggested an X-ray orbital period of 28.47 d.

\begin{figure}
\scalebox{0.46}{\includegraphics{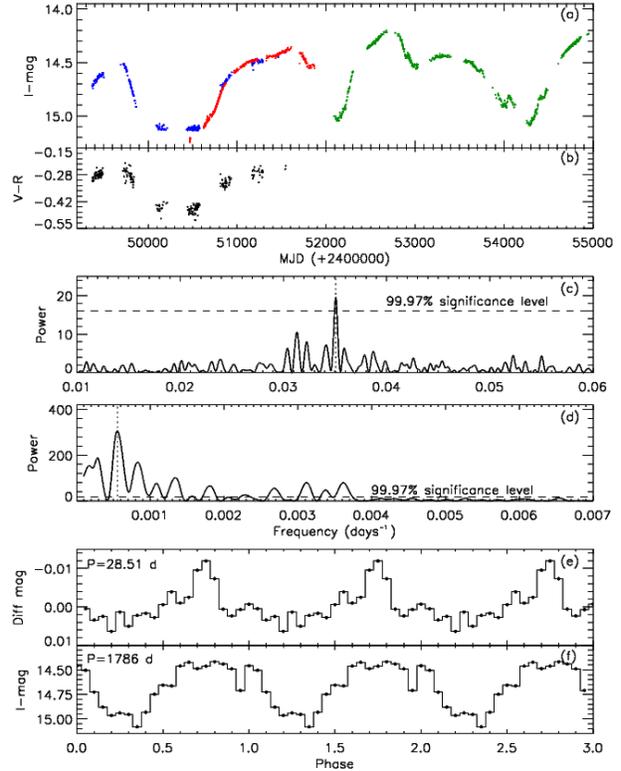}}
\caption{SXP8.88. (a): MACHO $\mathcal{R}$-band (blue), OGLE II (red) and OGLE III (green) light curves. (b): MACHO colour variation ($\mathcal{V}-\mathcal{R}$). (c) and (d): power spectrum of the detrended and combined data with highest peaks at P=28.51 d and P=1786 d. (e) and (f): Light curve folded on periods of 28.51 d and 1786 d respectively.}
\label{sxp8.9lc} 
\end{figure}

Given the large amplitude long-term variation present in the combined MACHO and OGLE lightcurves of SXP8.88 (Figure~\ref{sxp8.9lc}a), it is not surprising that the power spectrum has significant low-frequency power with a peak at 1786$\pm$32 d (Figure~\ref{sxp8.9lc}d).

We have only selected and detrended the data during the bright phase (MJD 51000-52000 and MJD 52400-53400). The power spectrum of our detrended OGLE light curve shows a significant peak at a period of 28.51$\pm$0.01 d which is very close to the X-ray period reported by \citet{corbet04a} (Figure~\ref{sxp8.9lc}c). The folded light curve on this 28.51 d period is shown in Figure~\ref{sxp8.9lc}e, however, the amplitude is very weak.

\subsection{SXP9.13 (AX J0049-732)}

This source was found to be a 9.1321 s X-ray pulsar by \citet{imanishi98}. \citet{schmidtke04} proposed RX J0049.5-7310 as a ROSAT counterpart of this ASCA pulsar. However, \citet{coe05} pointed out that RX J0049.5-7310 lies well outside the revised X-ray error circle presented in \citet{ueno01} and proposed RX J0049.2-7311 to be the more likely ROSAT counterpart, and identified its optical counterpart. A period of 40.17 d was reported by \citet{edge05a}, however \citet{galache08} suggested an X-ray orbital period of 77.2$\pm$ 0.3 d.

\begin{figure}
\scalebox{0.46}{\includegraphics{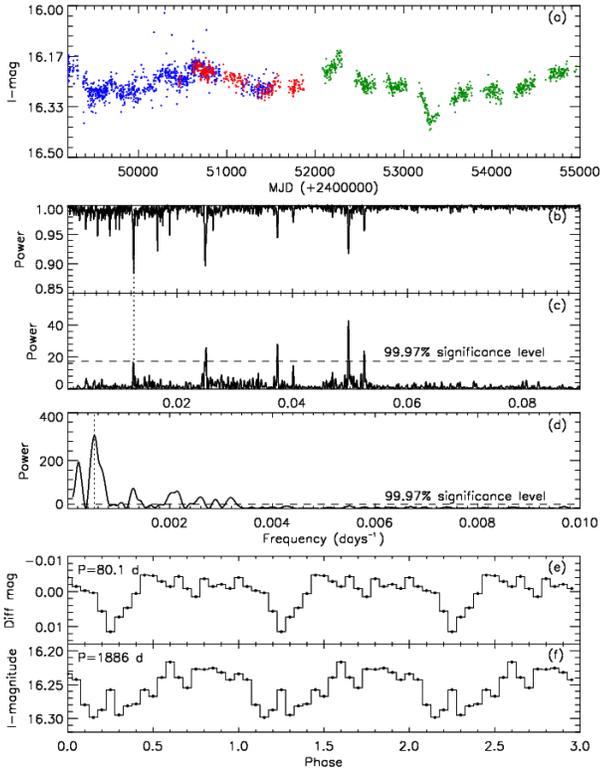}}
\caption{SXP9.13. (a): MACHO $\mathcal{R}$-band (blue), OGLE II (red) and OGLE III (green) light curves. (b) and (c): Power spectra (PDM and L-S) of the detrended light curve. (d): Power spectrum of the combined light curve. (e) and (f): Light curve folded on periods of 80.1 d and 1886 d respectively.}
\label{sxp9.13lc} 
\end{figure}

We detrended the OGLE II and OGLE III data by subtracting a linear fit to the long-term trends. The periodograms revealed a peak at 80.1$\pm$0.06 d in both PDM and L-S periodograms, plus its harmonics at 20, 26.6 and 40 d (figure~\ref{sxp9.13lc}b and \ref{sxp9.13lc}c). This period is very similar to the X-ray orbital period found by \citet{galache08} and twice the optical period reported by \citet{edge05a}. The folded light curve on the period of 80.1 d is shown in Figure~\ref{sxp9.13lc}e. The shape of the light curve is unusual with a dip of low amplitude.

The power spectrum of the combined long-term light curves has a strong peak with a superorbital period of 1886$\pm$35 d (Figure~\ref{sxp9.13lc}d). The folded light curve on this longer period is shown in Figure~\ref{sxp9.13lc}f.

\subsection{SXP15.3 (RX J0052.1-7319)}

The 15.3 s pulsations from RX J0052.1-7319 were discovered by \citet{lamb99}. Subsequently, \citet{israel99} proposed its optical counterpart to be a B-type star, which was confirmed spectroscopically by \citet{covino01}.

\begin{figure}
\scalebox{0.46}{\includegraphics{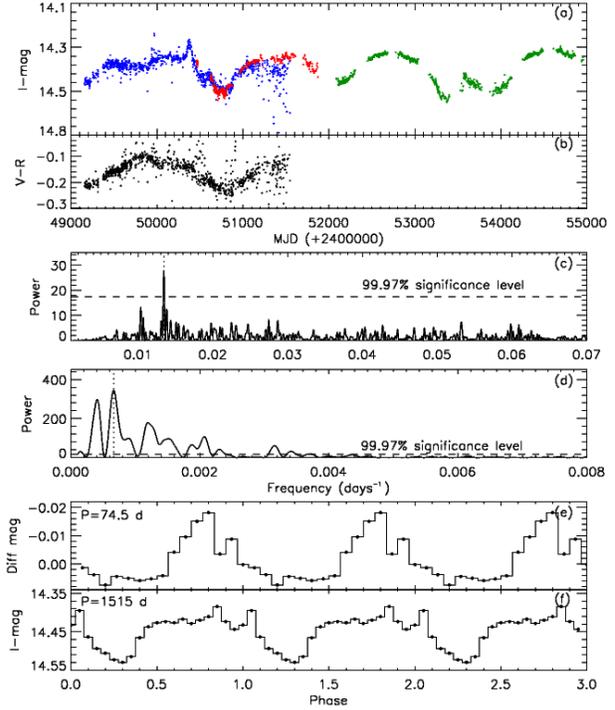}}
\caption{SXP15.3. (a) MACHO $\mathcal{R}$-band (blue), OGLE II (red) and OGLE III (green) light curves. (b) MACHO colour variation ($\mathcal{V}-\mathcal{R}$). (c) and (d): Power spectrum of the detrended and combined light curve showing peaks at P= 74.5 d and P= 1515 d. (e) and (f): Light curve folded on periods of 74.5 d and 1515 d respectively.}
\label{sxp15.3lc} 
\end{figure}

The combined MACHO and OGLE light curves (Figure~\ref{sxp15.3lc}d) display substantial, long-term, quasi-periodic variations of 1515$\pm$23 d, and the light curve folded on this superorbital period is shown in Figure~\ref{sxp15.3lc}f. The power spectrum of the detrended OGLE II and OGLE III light curves also show a peak at 74.5$\pm$0.05 d (Figure~\ref{sxp15.3lc}c) which is consistent with the orbital period of 75.1 d found by \citet{edge05a}. Figure~\ref{sxp15.3lc}e shows the detrended light curve folded on this period. The 74.5 d modulation is present throughout the 1515 d cycle.

\subsection{SXP18.3 (XMMJU004911.4-724939)}
\begin{figure}
\scalebox{0.46}{\includegraphics{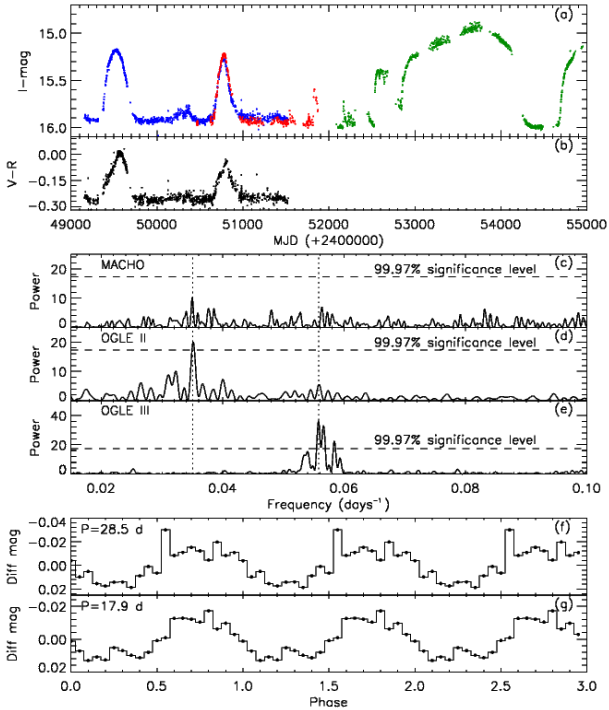}}
\caption{SXP18.3. (a): MACHO $\mathcal{R}$-band (blue), OGLE II (red) and OGLE III (green) light curves. (b): MACHO colour variation ($\mathcal{V}-\mathcal{R}$). (c),(d),(e): Power spectrum of the detrended MACHO, OGLE II and second part of OGLE III data, respectively. (f),(g): The light curves folded on periods of 28.5 d and 17.9 d respectively.}
\label{sxp18.3lc} 
\end{figure}

This X-ray source was discovered to be an X-ray pulsar by \citet{corbet03b} with a period of 18.37$\pm$0.01 s. The XMM-Newton position \citep{eger08} allowed its optical counterpart to be identified as a V=16 star \citep{zaritsky02}. A presumed orbital period of 17.79 d was found using both optical \citep{coe08,schurch09}, and X-ray observations \citep{galache08}. The temporal analysis finds no significant long-term periodicities in the combined light curves, although there are extreme excursions ($\geq 1~mag$) on timescales of $\sim$ 2000 d (Figure~\ref{sxp18.3lc}a). In Figure~\ref{sxp18.3lc}b, it is clear that the source becomes redder when it brightens.

We have detrended the MACHO and OGLE datasets and examined their temporal properties in sections with surprising results. The light curves from the MACHO, OGLE II, and the first year of OGLE III data show a clear variation of 28.5$\pm$0.01 d, which appears as a significant peak in its power spectrum (Figure~\ref{sxp18.3lc}c and \ref{sxp18.3lc}d). The folded light curve on this period (Figure~\ref{sxp18.3lc}f) shows a clear sinusoidal variation. However, for the OGLE III data (from MDJ 53000), its power spectrum shows a significant peak at a period of 17.9$\pm$0.01 d (Figure~\ref{sxp18.3lc}e), which is very similar to the reported X-ray 17.73 d  period by \citet{galache08}. The folded light curve on this period is shown in Figure~\ref{sxp18.3lc}g.

\begin{figure}
\scalebox{0.46}{\includegraphics{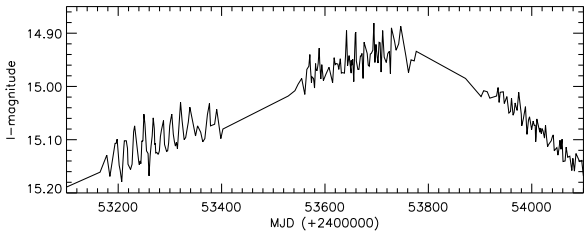}}
\caption{SXP18.3. Blow-up of the OGLE III light curve showing the 17.9 d modulation.}
\label{sxp18.3lc2} 
\end{figure}

The optical modulation of 28.5 d only appears in the MACHO, OGLE II, and first year of the OGLE III data where I$\sim$ 15.9. But, when the source becomes brighter (I$\leq$ 15.2) the modulation at 17.9 d starts to be seen in the light curve (Figure~\ref{sxp18.3lc2}).

We note that the 28.5 d modulation disappears during and after the long type II outburst which lasts for about 1500 d even as the brightness of the source is returning to its quiescent state (I$\sim$ 15.9). Both modulations are of low amplitude.

\subsection{SXP22.1 (RX J0117.6-7330)}

This was first discovered with ROSAT by \citet{clark96} as an X-ray transient. Then, \citet{charles96} identified its optical companion as a 14.2 mag OB star which was confirmed by \citet{coe98} as a Be(B1-2) star. The 22 s pulsation was found by \citet{macomb99}.

\begin{figure}
\scalebox{0.46}{\includegraphics{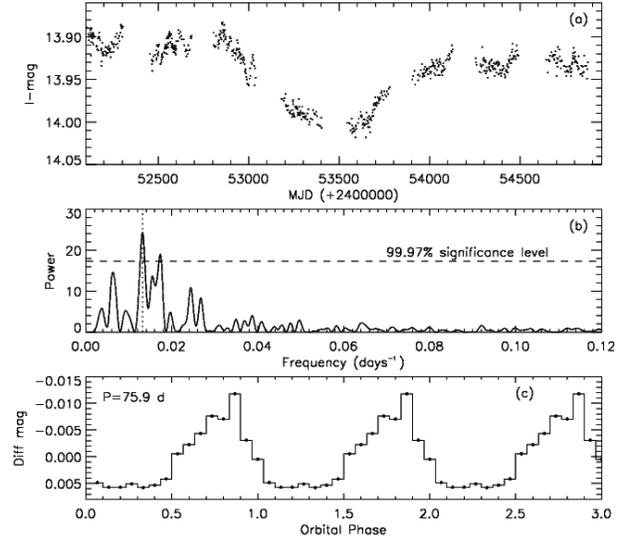}}
\caption{SXP22.1. (a): OGLE III light curve. (b): The power spectrum of the detrended OGLE III light curve showing the peak at P= 75.9 d. (c): The light curve folded on the orbital period of 75.9 d.}
\label{sxp22.1lc} 
\end{figure}

After detrending the data by subtracting a linear fit, we found a peak at a period of 75.9$\pm$0.06 d in its power spectrum (Figure~\ref{sxp22.1lc}b). The 75.9 d orbital outbursts appear to be strong when the source is brighter, and very weak at optical minima (Figure~\ref{sxp22.1lc}a). The folded light curve on this period is shown in Figure~\ref{sxp22.1lc}c. The long-term light curve is clearly variable on a timescale of $\sim$ 2000 d, which is comparable to the duration of the dataset.

\subsection{SXP25.5 (XMMU J004814.1-731003)}

\citet{lamb02} found the 25.55 s pulsation of this X-ray source. The accuracy of its X-ray position with XMM allowed its optical companion to be identified as a V=15.5 star by \citet{haberl08b}.

\begin{figure}
\scalebox{0.46}{\includegraphics{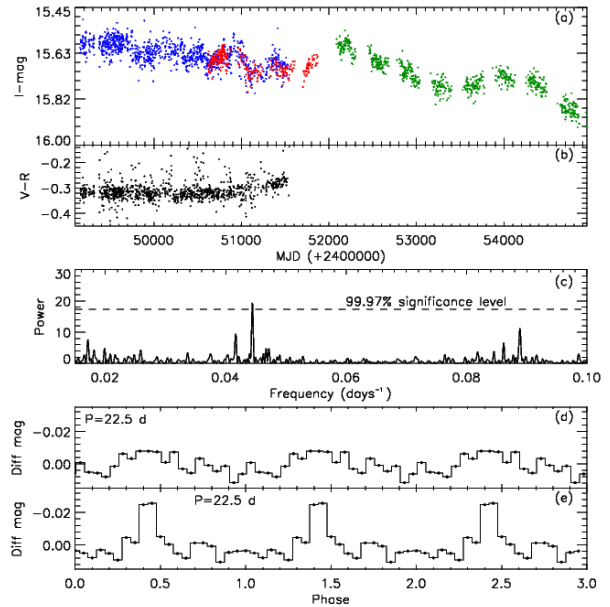}}
\caption{SXP25.5. (a): MACHO $\mathcal{R}$-band (blue), OGLE II (red) and OGLE III (green) light curves. (b): MACHO colour variation ($\mathcal{V}-\mathcal{R}$). (c): Power spectrum of the OGLE III light curve. (d),(e): First (top) and second (bottom) part of the OGLE III light curve folded on the orbital period of 22.5 d.}
\label{sxp25.5lc} 
\end{figure}

Analysis of the detrended OGLE III data reveals a significant peak at a period of 22.5$\pm$0.005 d (Figure~\ref{sxp25.5lc}c). The 22.5 d period appears to be caused by several outbursts in the OGLE III data, and the folded light curve on this period shows a typical BeX outburst behaviour (Figure~\ref{sxp25.5lc}e).

We subdivided the OGLE III data into two parts, and then we folded it on the above period. The amplitude of the outbursts appear to vary as a function of brightness, they are weak when the source is brighter (first part of OGLE III data) and strong as the source becomes fainter (second part) (Figure~\ref{sxp25.5lc}d and ~\ref{sxp25.5lc}e). Furthermore, the colour variation in Figure~\ref{sxp25.5lc}b shows a behaviour similar to that seen in A0538-66, where the source is bluer when brighter.

\subsection{SXP31.0 (XTE J0111.2-7317)}

SXP31.0 was first discovered with ASCA by \citet{chakrabarty98} as a 31s X-ray pulsar. Its optical companion was identified as a B0-B2 star by \citet{coe00}. \citet{schmidtke06} reported an orbital period of P = 90.4 d using OGLE III data.

\begin{figure}
\scalebox{0.46}{\includegraphics{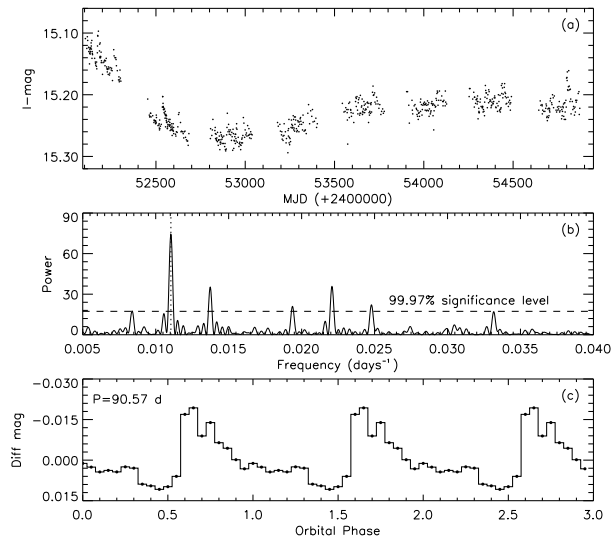}}
\caption{SXP31.0. (a) OGLE III light curve. (b): Power spectrum of the OGLE III light curve showing the peak at 90.57 d. (c): The OGLE III light curve folded on the orbital period of 90.57 d.}
\label{sxp31.0lc} 
\end{figure}

After detrending the OGLE III light curve, we find a period of 90.57$\pm$0.08 d, very similar to that already reported (Figure~\ref{sxp31.0lc}b). The light curve folded on this period is shown in Figure~\ref{sxp31.0lc}c. This classic BeX profile with outbursts every 90.57 d indicates that it is certainly the orbital period of SXP31.0. It is present even when the source is at optical minimum. Furthermore, the amplitude of the outbursts varies, being very strong during the two first years and one very strong outburst near the end of the OGLE III observations.

We do not find any long-term periodicity from the lightcurve (Figure~\ref{sxp31.0lc}a), although it is clearly variable on a timescale of $\sim$2000 d.

\subsection{SXP34.1 (CXOU J005527.9-721058)}

The 34 s pulsation from this source was found by \citet{edge04} using Chandra data. The precise X-ray position led to an optical identification as a B-type star \citep{coe05}. The orbital period of SXP34.1 has not yet been reported.

\begin{figure}
\scalebox{0.92}{\includegraphics{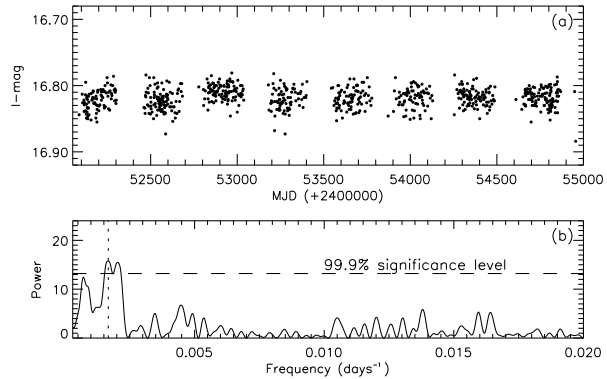}}
\caption{SXP34.1. (a): OGLE III light curve. (b) Power spectrum of the OGLE III light curve showing a significant peak at 598 d.}
\label{sxp34.1lc} 
\end{figure}

The brightness of this BeX source is remarkably constant at I$\sim$ 16.82 during all OGLE III observations (Figure~\ref{sxp34.1lc}a). However, its power spectrum shows a marginally significant peak at 598$\pm$3.6 d (Figure~\ref{sxp34.1lc}b) with a very low amplitude.

However, we note that an orbital period of 598 d is too long for a 34 s BeX source based on the Corbet pulse/orbital period relationship.

\subsection{SXP46.6 (1WGA J0053.8-7226)}

\begin{figure}
\scalebox{0.46}{\includegraphics{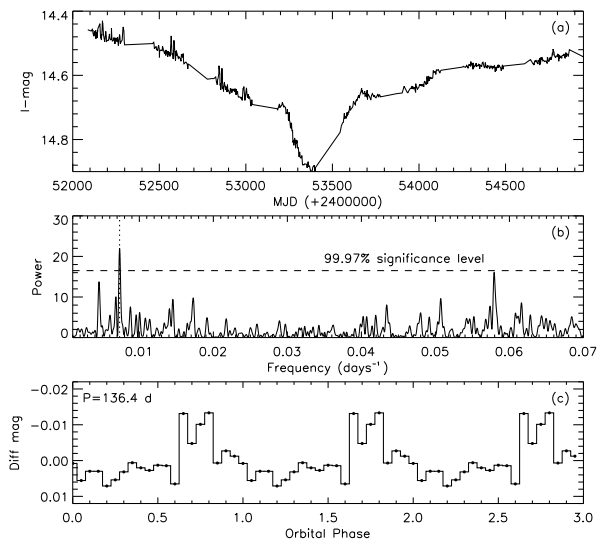}}
\caption{SXP46.6. (a): OGLE III light curve. (b): Power spectrum of the OGLE III light curve showing peaks at 136.4 d and 17.2 d. (c): The OGLE III light curve folded on 136.4 d.}
\label{sxp46.6lc} 
\end{figure}

\citet{marshall97} initially reported a pulse period of 92 s from this source, but further analysis of the ASCA observations by \citet{corbet98} revealed the period to be 46.63 $\pm$ 0.04 s. \citet{mcgowan08} confirmed its optical counterpart to be Star B from \citet{buckley01} with an optical orbital period of 136 $\pm$ 2 d (based on the data of Figure~\ref{sxp46.6lc}a), consistent with the X-ray period found by \citet{galache08}.

The light curve of SXP46.6 shows not only one but two peaks for each periastron passage of the X-ray pulsar. The power spectrum of the detrended light curve reveals significant peaks at 136.4$\pm$0.2 and 17.2$\pm$0.003 d (Figure~\ref{sxp46.6lc}b). The light curve folded on 136.4 d is typical of BeX orbital behaviour, which is shown in Figure~\ref{sxp46.6lc}c. The 17.2 d appears to be the time interval between the two peaks in each orbital cycle. 

We also find that the amplitude of the outbursts appears to become weaker as the source fades (Figure~\ref{sxp46.6lc}a). The OGLE III light curves show a large amplitude, long-term variation of about half a magnitude on timescales $\sim$ 2000 d, comparable to the duration of the dataset.

\subsection{SXP59.0 (XTE J0055-724)}

This source was reported as a 59.0$\pm$0.2 s X-ray pulsar by \citet{marshall98}. Its optical companion is identified as MA[93]810 and MACHO 207.16259.23. \citet{schmidtke05b} reported an orbital period of 60.2 d in the OGLE II data. However, \citet{galache08} suggested an X-ray orbital period of 122.1$\pm$0.38 d in their RXTE data.

\begin{figure}
\scalebox{0.46}{\includegraphics{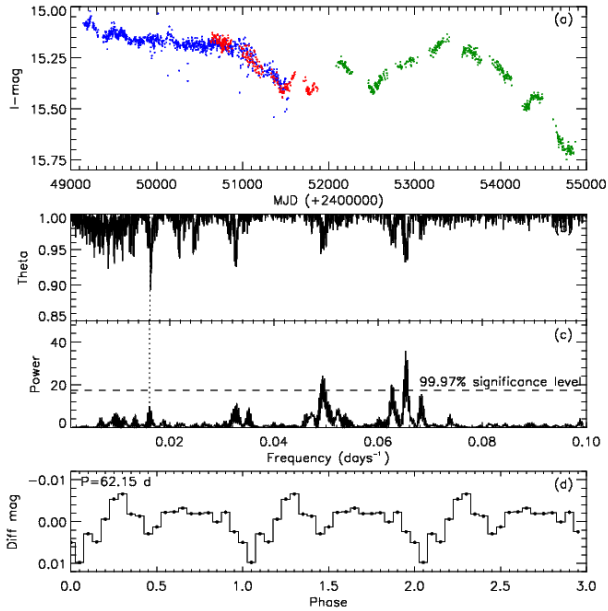}}
\caption{SXP59.0. (a): MACHO $\mathcal{R}$-band (blue), OGLE II (red) and OGLE III (green) light curves. (b): Power spectrum of the combined data showing peaks at 62.15 d period, and its harmonics at 15, 20, and 30 d. (c): Detrended combined light curve folded on the 62.15 d period.}
\label{sxp59.0lc} 
\end{figure}

The power spectrum (PDM and L-S ) of our detrended MACHO and OGLE combined lightcurve has a significant peak at 62.15$\pm$0.04 d, consistent with that previously reported, and its harmonics at 15, 20, and 30 d (Figure~\ref{sxp59.0lc}b, and \ref{sxp59.0lc}c). We note that \citet{galache08} found an X-ray period of 122 d which is about twice our value. The light curve folded on 62.15 d is shown in Figure~\ref{sxp59.0lc}d. The long-term light curve exhibits a large amplitude ($>$ 0.5 mag) variability on a timescale of $\sim$3000 d.

\subsection{SXP74.7 (AX J0049-729)}

The 74.7 s pulsation from AX J0049-729 was discovered by \citet{yokogawa98}. \citet{schmidtke05b} reported an orbital period of 33.4$\pm$0.4 d in the MACHO lightcurve of this source.

\begin{figure}
\scalebox{0.46}{\includegraphics{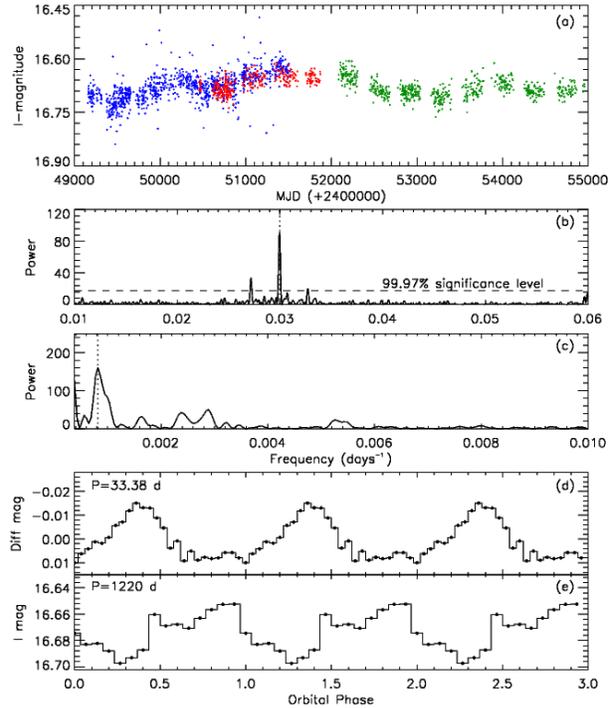}}
\caption{SXP74.7.  (a): MACHO $\mathcal{R}$-band (blue), OGLE II (red) and OGLE III (green) light curves. (b) and (c): Power spectrum of the detrended and combined light curve showing peaks at 33.38 d and 1220 d. (d) and (e): The light curves folded on 33.38 d and 1220 d period.}
\label{sxp74.7lc} 
\end{figure}

The power spectrum of the detrended light curve of the complete MACHO and OGLE observations shows a very significant peak at a period of 33.38 $\pm$ 0.01 d (Figure~\ref{sxp74.7lc}b), similar to that already reported. The light curve folded on this period is shown in Figure~\ref{sxp74.7lc}d.

The power spectrum of the long-term light curve shows a peak at 1220$\pm$64 d (Figure~\ref{sxp74.7lc}c), and the light curve folded on this period is shown in Figure~\ref{sxp74.7lc}e.

\subsection{SXP82.4 (XTE J0052-725)}

\citet{corbet02} first observed the pulse period of 82.4s for this X-ray pulsar with RXTE. Its position was determined from Chandra data by \citet{edge03b}.

\begin{figure}
\scalebox{0.46}{\includegraphics{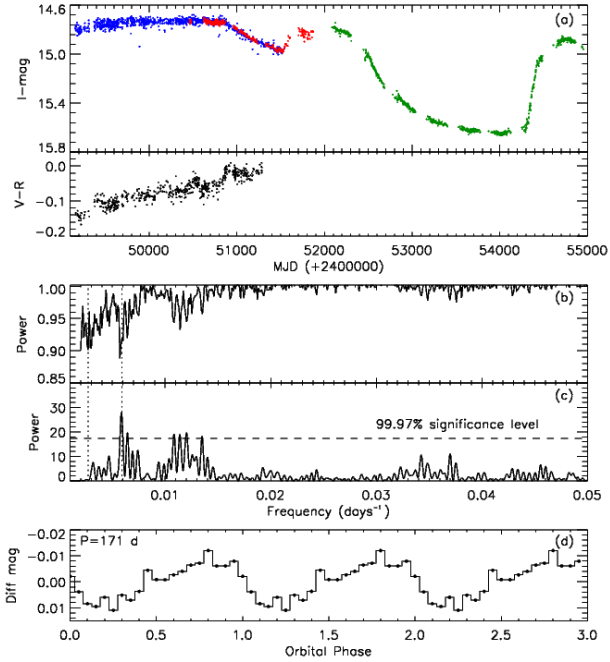}}
\caption{SXP82.4. (a): MACHO $\mathcal{R}$-band (blue), OGLE II (red) and OGLE III (green) light curves. (b) and (c): Power spectra (PDM and L-S) of the OGLE III light curve showing a significant peak at 171 d, the vertical dashed line represents the previously reported period. (d): The OGLE III light curve folded on the 171 d period.}
\label{sxp82.4lc}
\end{figure}

Its long-term light curve appears to be constant until MJD$\sim$50800, following which it shows a dramatic variation on timescales $\sim$ 2000 d, and declining by $\ge$ 1 mag (see Figure~\ref{sxp82.4lc}a).

\citet{schmidtke05b} have studied the MACHO dataset but were unable to find any signature of its orbital period. However, \citet{galache08} reported an X-ray period of $\sim$362.3 d, and derived the X-ray outburst ephemeris.

We have split the OGLE III datasets into two parts. The power spectra (PDM and L-S) of both parts show a significant peak at 171$\pm$0.3 d, which is less than half the reported X-ray period (Figure~\ref{sxp82.4lc}b). The PDM shows a peak at 370.2 d, very similar to the previously reported X-ray period, but it may be caused by the annual sampling of the data. The light curve folded on the 171 d period is shown in Figure~\ref{sxp82.4lc}d.

\subsection{SXP91.1 (RX J0051.3-7216)}

The 91.12 s pulse period of this X-ray pulsar was reported by \citet{corbet98}. \citet{stevens99} proposed its optical counterpart as an H$\alpha$ emitting object MA[93]413 \citep{schmidtke04}. An orbital period of 88.25 d was found by \citet{schmidtke04} in the MACHO light curve. 

\begin{figure}
\scalebox{0.46}{\includegraphics{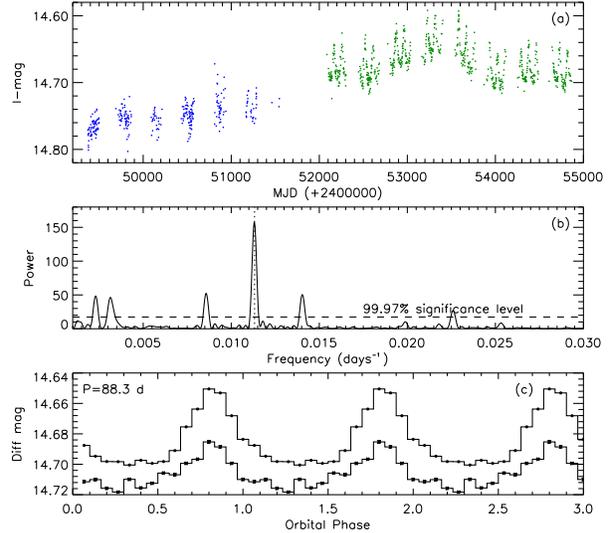}}
\caption{SXP91.1. (a): MACHO $\mathcal{R}$-band (blue), and OGLE III (green) light curves. (b): Power spectrum of the combined detrended light curves showing a peak at 88.3 d. (c): OGLE III and MACHO light curves folded on 88.3 d.}
\label{sxp91.1lc} 
\end{figure}

The power spectrum of the detrended light curves reveals a very significant peak at 88.3$\pm$0.1 d (Figure~\ref{sxp91.1lc}b). The folded light curve on the 88.3 d period is shown in Figure~\ref{sxp91.1lc}c. The combined long-term light curve varies on timescales of $\sim$4000 d. Moreover, the amplitude of the 88.3 d outbursts was clearly variable throughout this 4000 d cycle, appearing stronger when the source brightens (Figure~\ref{sxp91.1lc}a).

\subsection{SXP101 (AX J0057.4-7325)}

This variable X-ray source was found to be a 101.45 s X-ray pulsar by \citet{yokogawa00a}. \citet{mcgowan07} confirmed its optical counterpart as a V = 14.9 star with a period (presumed to be orbital) of 21.9 d in both MACHO and OGLE III light curves.

\begin{figure}
\scalebox{1}{\includegraphics{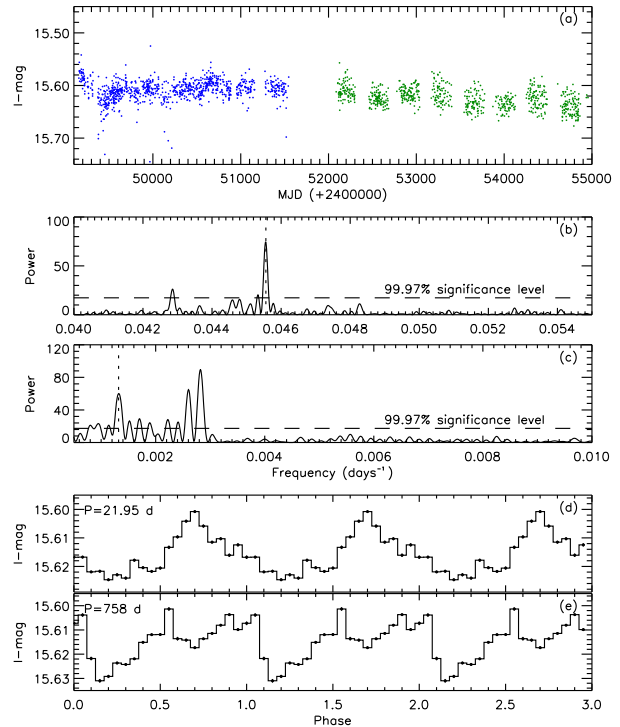}}
\caption{SXP101. (a): MACHO $\mathcal{R}$-band (blue), and OGLE III (green) light curves. (b) and (c): Power spectrum of the combined MACHO and OGLE light curves showing a significant peak at 21.95 d and 758 d. (d) and (e): The combined light curve folded on 21.95 d and 758 d period, respectively.}
\label{sxp101lc} 
\end{figure}

We detrended the complete MACHO and OGLE III datasets and found a peak consistent with the previously reported period at 21.95$\pm$0.004 d (Figure~\ref{sxp101lc}b). The light curve folded on this period is shown in Figure~\ref{sxp101lc}d. The power spectrum of the combined light curve of SXP101 also shows peaks at 758$\pm$6 d and 355 d (Figure~\ref{sxp101lc}c). The peak at 355 d is due to the data sampling. The folded light curve on the 758 d period is shown in Figure~\ref{sxp101lc}e.

\subsection{SXP138 (CXOU J005323.8-722715)}

The 138.04 s pulse period of this source was reported by \citet{edge04} using archival Chandra data. It was identified with MA[93]667 and found to have a period of $\sim$ 125 d in MACHO data \citep{edge05a}. This was confirmed by \citet{schmidtke06} who found a period of 122-123 d.

\begin{figure}
\scalebox{1}{\includegraphics{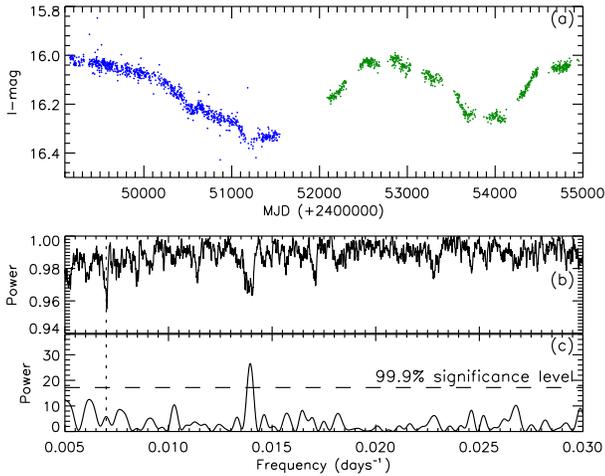}}
\caption{SXP138. (a): MACHO $\mathcal{R}$-band (blue), and OGLE III (green) light curves. (b) and (c): Power spectrum (PDM and L-S) of the detrended light curve with a peak at P=143 d.}
\label{sxp138lc} 
\end{figure}

The L-S power spectrum of the detrended MACHO and OGLE III data shows a significant peak at 71.8$\pm$0.1 d, however the deepest peak in the PDM occurs at twice this value, 143$\pm$0.2 d (Figure~\ref{sxp138lc}b, and ~\ref{sxp138lc}c). We found a very weak peak at 125 d, but in the MACHO data only.

The long-term light curve shows a large variation of about 0.4 mag with a period of 2700 d.

\subsection{SXP140 (XMMU J005605.2-722200)}

\citet{sasaki03} discovered this 140 s pulsar in XMM-Newton observations, whose accurate X-ray position clearly identifies it with MA[93]904. \citet{schmidtke06} found a 197$\pm$5 d period in a segment of MACHO light curve (MJD 49650-50500) which they suggested as the orbital period.

\begin{figure}
\scalebox{0.46}{\includegraphics{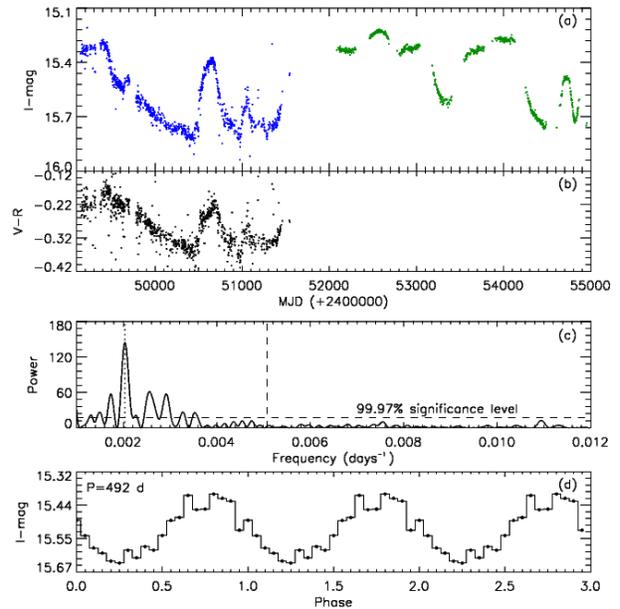}}
\caption{SXP140. (a) MACHO $\mathcal{R}$-band (blue), and OGLE III (green) light curves. (b): MACHO colour variation $\mathcal{V}$- $\mathcal{R}$. (c): Power spectrum of the combined light curve with a highest peak at 492 d, the vertical dash line represents the frequency of the previously reported period. (d): The combined light curves folded on the period of 492 d.}
\label{sxp140lc} 
\end{figure}

The light curves show large amplitude, long-term variations of half a magnitude, but the power spectrum of the OGLE III light curve reveals nothing significant at the previously reported orbital period. However, it does show a very strong peak at 492$\pm$2.4 d (Figure~\ref{sxp140lc}c), and the combined light curve folded on this period is shown in Figure~\ref{sxp140lc}d.

Figure~\ref{sxp140lc}b shows the colour variation MACHO $\mathcal{V}-\mathcal{R}$ which correlates closely with the flux modulation. A lag between the MACHO colour and brightness variation is also seen at its optical maximum.

\subsection{SXP172 (RX J0051.9-7311)}

The optical counterpart of this source was identified as a Be star by \citet{cowley97}, and is also known as MA[93]504. Subsequently, \citet{yokogawa00b} found an X-ray pulsation with a period of 172.4 s, which was refined by \citet{haberlandpietsch04} to 172.21$\pm$0.13 s. \citet{laycock05} reported a 67$\pm$5 d period based on several X-ray outbursts. An orbital period of 69.9$\pm$0.6 d was proposed by \citet{schmidtke06} using only the first two years of the OGLE II data (which they named Segment A).

\begin{figure}
\scalebox{0.46}{\includegraphics{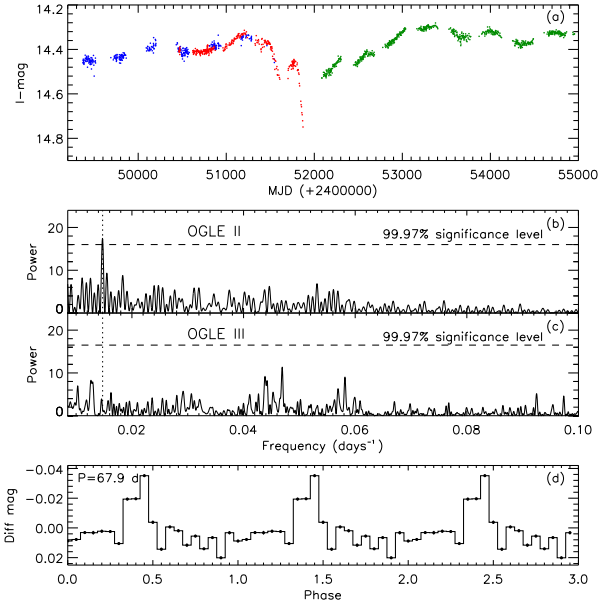}}
\caption{SXP172. (a): MACHO $\mathcal{R}$-band (blue), OGLE II (red) and OGLE III (green) light curves. (b) and (c): Power spectrum of the detrended OGLE II (peak at P=67.9 d) and OGLE III light curves, the vertical dotted line represents the frequency of the previously reported 69.9 d orbital period. (d): The folded OGLE II light curve on the 67.9 d period}
\label{sxp172lc} 
\end{figure}

Our power spectrum of the detrended OGLE II light curve shows a peak at 67.9$\pm$0.04 d (Figure~\ref{sxp172lc}b), close to that previously reported. However, the OGLE III power spectrum (Figure~\ref{sxp172lc}c) does not show a peak at this period. The folded OGLE II light curve on the 67.9 d period is shown in Figure~\ref{sxp172lc}d. The long-term light curve shows a large amplitude variation on a timescale of $\sim$ 2000 d (Figure~\ref{sxp172lc}a).

\subsection{SXP202A (1XMMU J005921.0-722317)}

\citet{majid04} discovered the 202 s pulsation of this X-ray source in a number of archival XMM-Newton observations. The optical counterpart was identified as MACHO 207.16545.12.

\begin{figure}
\scalebox{0.46}{\includegraphics{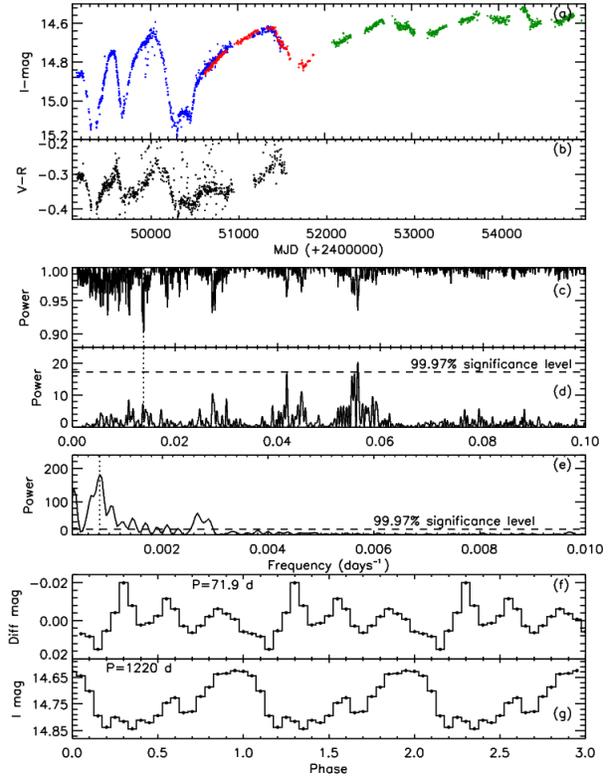}}
\caption{SXP202A. (a): MACHO $\mathcal{R}$-band (blue), OGLE II (red), and OGLE III (green) light curves. (b): MACHO colour variation $\mathcal{V}$- $\mathcal{R}$'. (c) and (d): PDM and L-S periodogram of the detrended light curve showing a peak at 71.9 d and its harmonics. (e): Power spectrum of the combined light curve with a peak at P=1220 d. (f)and (g): The light curves folded on the 71.9 d, and 1220 d periods.}
\label{sxp202Alc} 
\end{figure}

The light curve is highly variable by $\ge$ 0.5 mag, and the power spectrum of the combined light curves revealed strong low frequency power with a peak at P=1220$\pm$61 days (Figure~\ref{sxp202Alc}e). The folded lightcurve on this long-term period is shown in Figure~\ref{sxp202Alc}g. In Figure~\ref{sxp202Alc}b, we can see that the optical maxima and the peaks of the colour variation (MACHO $\mathcal{V}$- $\mathcal{R}$) are not synchronized.

Using the RXTE/PCA observation, \citet{galache08} reported that they did not find any clear orbital period in the data for SXP202. However, they noted that the time interval between the six X-ray maxima detected since MJD 53000 is about $\sim$91 d. After detrending the optical data from MACHO and OGLE observations, we found a peak at 71.9$\pm$0.05 d and several peaks at its harmonics (36 d, 23 d, and 18 d) in both PDM and L-S power spectra (Figure~\ref{sxp202Alc}c and \ref{sxp202Alc}d). The detrended light curve folded on this period is complex (Figure~\ref{sxp202Alc}f).

\subsection{SXP202B (XMMU J005929.0-723703)}

XMMU J005929.0-723703 was reported as a new X-ray pulsar with a pulse period of 202 s by \citet{haberl08a}. They identified its optical companion with a V=14.9 Be star (also MA[93]1147), and classified it as a B0-5(III)e in the 2dF survey of the SMC.

\begin{figure}
\scalebox{0.46}{\includegraphics{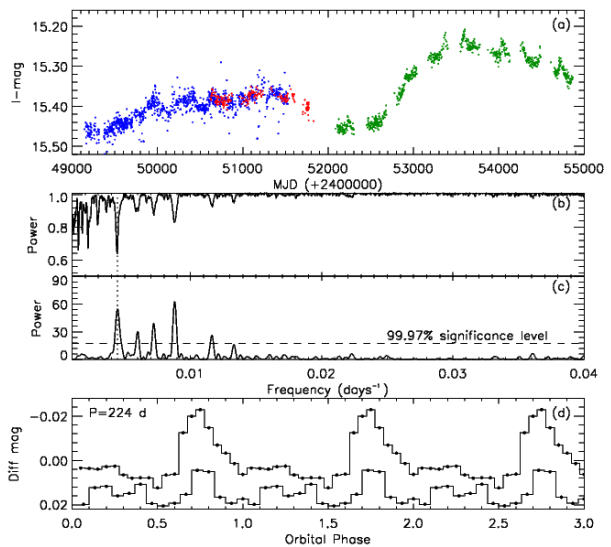}}
\caption{SXP202B. (a): MACHO (blue), OGLE II (red), and OGLE III (green) light curves. (b) and (c): Power spectra (PDM and LS) of the OGLE III light curve showing peak at 224 d. (d): The OGLE III light curve folded on 224 d period at optical maxima and minima.}
\label{sxp202Blc} 
\end{figure}

Its light curves show large amplitude ($\sim$ 0.5 mag) variations on a timescale of $\sim$ 3000 d (Figure~\ref{sxp202Blc}a). The power spectra (PDM and L-S) of the detrended light curve show peaks at P = 224$\pm$0.5 d and its harmonic 113 d (Figure~\ref{sxp202Blc}b and \ref{sxp202Blc}c). The highest peak in the PDM is at P= 224 d, which is consistent with the time interval between consecutive outbursts in Figure~\ref{sxp202Blc}a.

The 224 d modulation appears as a sequence of outbursts during the last five years of the OGLE III observations. Furthermore, the folded light curves on the period of 224 d shows a classic BeX outburst profile (Figure~\ref{sxp202Blc}d), which suggests that the 224 d period is orbital. This period is in good agreement with the \citet{corbet84} P(pulse)/P(orbit) relation.

The amplitude of outbursts is clearly variable (Figure~\ref{sxp202Blc}d) and appears larger when the source is brighter, and either lower or even absent when fainter (see Section~\ref{outbamp}, Figure~\ref{outburstvariation}).

\subsection{SXP264 (RX J0047.3-7312)}

The 264$\pm$1 s pulsations were discovered by \citet{ueno04}, whose accurate XMM-Newton location allowed its optical counterpart to be identified clearly as a Be star MA[93]172. It also appears in MACHO (MACHO 212.15792.77) and OGLE catalogues of variable stars. \citet{schmidtke05b} reported an orbital period of 49.1 d in OGLE data.

\begin{figure}
\scalebox{0.46}{\includegraphics{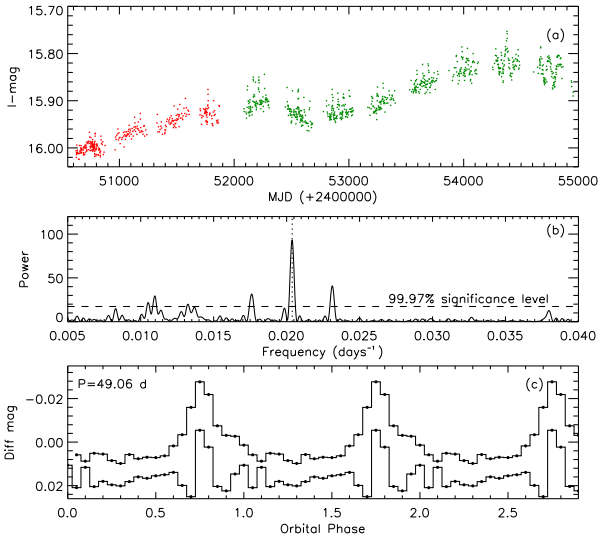}}
\caption{SXP264. (a): OGLE II, and OGLE III light curves. (b): Power spectrum of the detrended light curve showing a peak at 49.06 d. (c): OGLE II and OGLE III light curves folded on the 49.06 d orbital period.}
\label{sxp264lc} 
\end{figure}

Analysis of the detrended light curve reveals a significant peak at 49.06$\pm$0.02 d, consistent with that previously reported (Figure~\ref{sxp264lc}b). The light curve folded on this period is shown in Figure~\ref{sxp264lc}c. The combined light curve is clearly variable on a timescale of $\sim$2000 d with an increase in brightness over the last $\sim$ 5000 d (Figure~\ref{sxp264lc}a). The amplitude of the 49 d orbital outburst increases with its brightness.

\subsection{SXP280 (RX J0057.8-7202)}

\begin{figure}
\scalebox{0.46}{\includegraphics{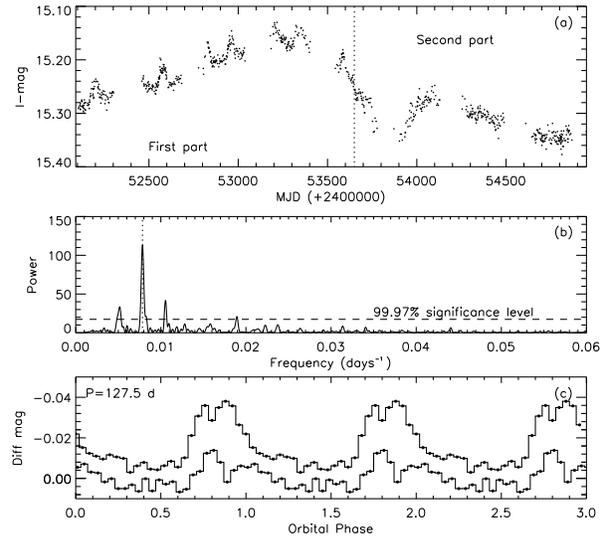}}
\caption{SXP280. (a): OGLE III light curve. (b): Power spectra of the detrended light curve with a peak at 127.5 d. (c): First and second half of the OGLE III light curve folded on the 127.5 d period.}
\label{sxp280lc} 
\end{figure}

The optical counterpart of this 280.4 s X-ray pulsar is identified as the Be star MA[93]1036. Its long-term light curve shows a clear series of outbursts every 127.3 d \citep{schmidtke06}. The power spectrum of the detrended light curve is dominated by a peak consistent with this at 127.5$\pm$0.2 d (Figure~\ref{sxp280lc}b).

We split the OGLE III data into two parts. Figure~\ref{sxp280lc}c represents these two parts of the OGLE III light curve folded on the 127.5 d period. The outburst amplitude of SXP280 also varies significantly ($\sim$ 0.25 mag) on a timescale of $\sim$ 2000 d.

\subsection{SXP293 (RX J0058.2-7231)}

\citet{corbet04b} detected the 292.9$\pm$0.4 s pulsations in RXTE observations of the SMC. The optical counterpart was identified as a Be star by \citet{schmidtke99}. \citet{galache08} suggested an orbital period of 151 d in their X-ray data from RXTE observation. However, both MACHO and OGLE light curves show clearly the orbital modulation with a period of P= 59.72 d, which appears as a series of very regular outbursts. 

\begin{figure}
\scalebox{0.46}{\includegraphics{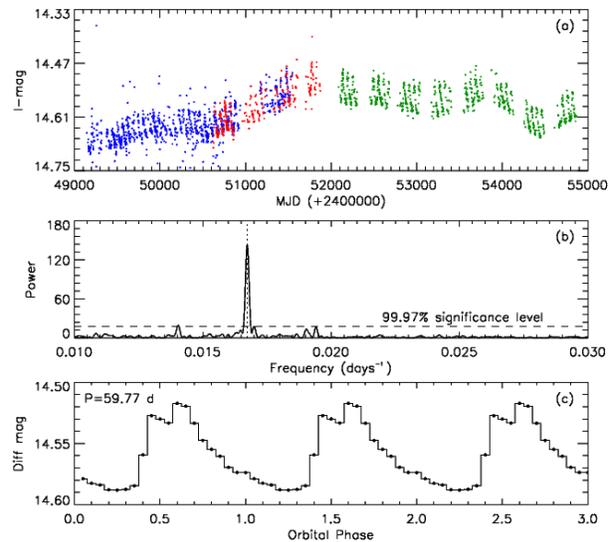}}
\caption{SXP293. (a): MACHO (blue), OGLE II (red), and OGLE III (green) light curves. (b): Power spectrum of the combined light curve showing a peak at 59.77 d period. c) The light curve folded on the 59.77 d period.}
\label{sxp293lc} 
\end{figure}

The combined light curves from MACHO and OGLE observations are shown in Figure~\ref{sxp293lc}a, and SXP293 shows a substantial increase in brightness until the end of the OGLE II observations. The power spectrum of the combined light curves shows a significant peak at a period of 59.77$\pm$0.03 d (Figure~\ref{sxp293lc}b), and the folded light curve on this period shows a double peaked classical BeX profile (Figure~\ref{sxp293lc}c). The long-term light curve is variable on a timescale of $\sim$2500 d

\subsection{SXP304 (RX J0101.0-7206)}

This 304.49 s X-ray pulsar was discovered in Chandra observations by \citet{macomb03}. Its X-ray position is coincident with the emission-line star MA[93]1240. \citet{schmidtke06} suggested a possible orbital period of 520$\pm$12 d in the MACHO data.

\begin{figure}
\scalebox{1}{\includegraphics{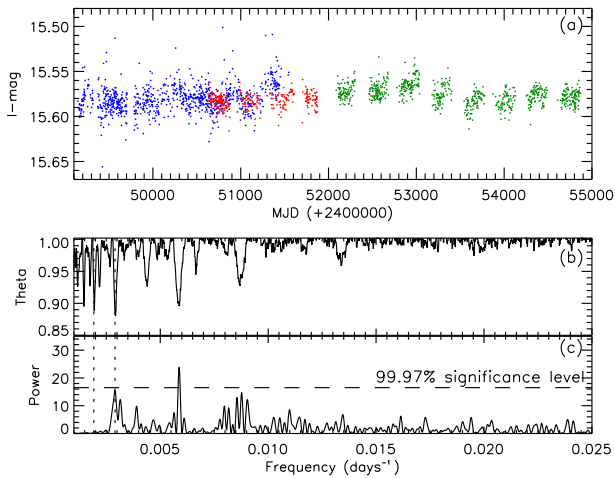}}
\caption{SXP304. (a): MACHO (blue), OGLE II (red), and OGLE III (green) light curves. (b) and (c): PDM and L-S periodograms of the combined light curves showing a peak at 344 d, the dashed lines represent the 344 d and the previously reported 520 d.}
\label{sxp304lc} 
\end{figure}

The power spectrum of the extended detrended dataset plotted in Figure~\ref{sxp304lc}c shows peaks at 344$\pm$1.2 d (deepest peak in the PDM) and its harmonic 170$\pm$0.3 d. The previously reported period 520 d appears as a peak in the PDM statistic of the combined data, but its power is lower than the 344 d period (Figure~\ref{sxp304lc}b).

\subsection{SXP327 (XMMU J005252.1)}

The pulsation of 327 s was discovered in the Chandra observations of the SMC by \citet{laycock08}. Its precise X-ray position allows the optical companion to be identified with object 207.16147.60 in the MACHO catalogue, and SMC101.4 25097 in the OGLE III catalogue. \citet{coe08} reported its orbital period to be 45.995 d using MACHO and OGLE III data. 

\begin{figure}
\scalebox{0.46}{\includegraphics{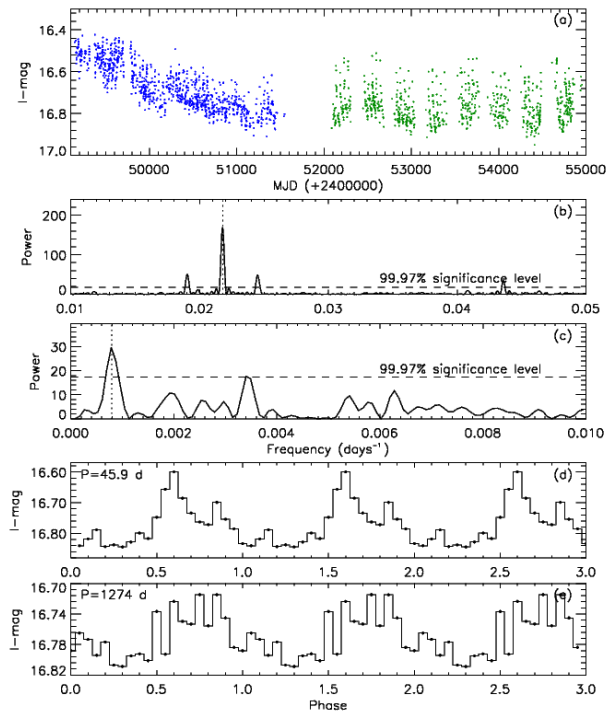}}
\caption{SXP327. (a): MACHO $\mathcal{R}$-band (blue), and OGLE III (green) light curves. (b) and (c): Power spectra of the combined light curve showing peaks at the presumed orbital period of 45.9 d, and 1274 d. (d) and (e): The combined light curves folded on 45.9 d and 1274 d respectively.}
\label{sxp327lc} 
\end{figure}

The MACHO light curve shows only a gradual fading, but the OGLE III light curve appears variable on longer timescales. The power spectrum of OGLE III data reveals peaks at 1274$\pm$143 d and the orbital period P= 45.9$\pm$0.2 d (Figure~\ref{sxp327lc}b and \ref{sxp327lc}c). The folded light curve on the orbital period (Figure~\ref{sxp327lc}d) shows a classic BeX profile, with an outburst amplitude of $\sim$ 0.25 mag. In addition, the orbital lightcurve shows a secondary maximum, which can arise due to misalignment between the circumstellar disc and the orbital plane of the neutron star.

\subsection{SXP348 (AX J0103-722)}

A 345.2 s pulsation was first detected in the BeppoSAX observation of AX J0103-722 in the SMC in 1998 by \citet{israel98}, then \citet{yokogawa98b} reported a pulse period of 348.9 s in the 1996 ASCA observation of the same source. The X-ray position from Chandra is coincident with a 14.8 mag Be star \citep{hughes94a} which is in the catalogue of \citet{meys93} as MA[93]1367. \citet{schmidtke06} suggested an orbital period of 93.9 d based on the OGLE II data.

\begin{figure}
\scalebox{0.46}{\includegraphics{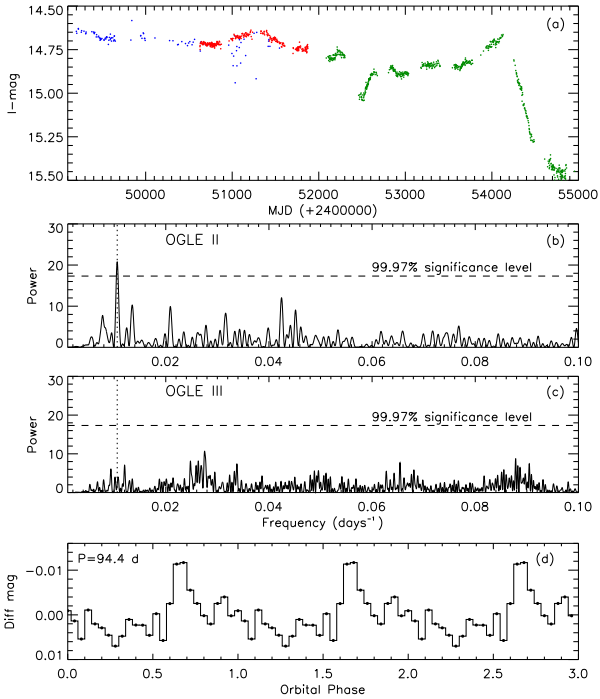}}
\caption{SXP348. (a): MACHO $\mathcal{R}$-band (blue), OGLE II, and OGLE III (green) light curves. (b): Power spectrum of the OGLE II light curve showing a significant peak at 94.4 d. (c): Power spectrum of the OGLE III light curve. (d): The folded OGLE II light curve on the 94.4 d period.}
\label{sxp345lc} 
\end{figure}

The power spectrum of our detrended OGLE II light curve shows a peak at 94.4$\pm$0.1 d (Figure~\ref{sxp345lc}b), very close to that previously reported. However, its power in the L-S periodogram of the OGLE III data is very low (Figure~\ref{sxp345lc}c). The OGLE II data folded on the orbital period of 94.4 d is shown in Figure~\ref{sxp345lc}d, but we note that the modulation amplitude is low. The source has faded dramatically recently (Figure~\ref{sxp345lc}a). The long-term lightcurve varies on a timescale of $\sim$ 2000 d (Figure~\ref{sxp345lc}a).

\subsection{SXP455 (RX J0101.3-7211)}

\citet{sasaki01} discovered the 455$\pm$2 s pulsations in XMM-Newton EPIC-PN data of RX J0101.3-7211. Its optical counterpart was suggested to be MA[93]1257 \citep{haberl00}, and optical observations by \citet{sasaki01} confirmed its Be nature. An orbital period of 74.7 d was found by \citet{schmidtke04} and confirmed by \citet{coe05}.

\begin{figure}
\scalebox{0.46}{\includegraphics{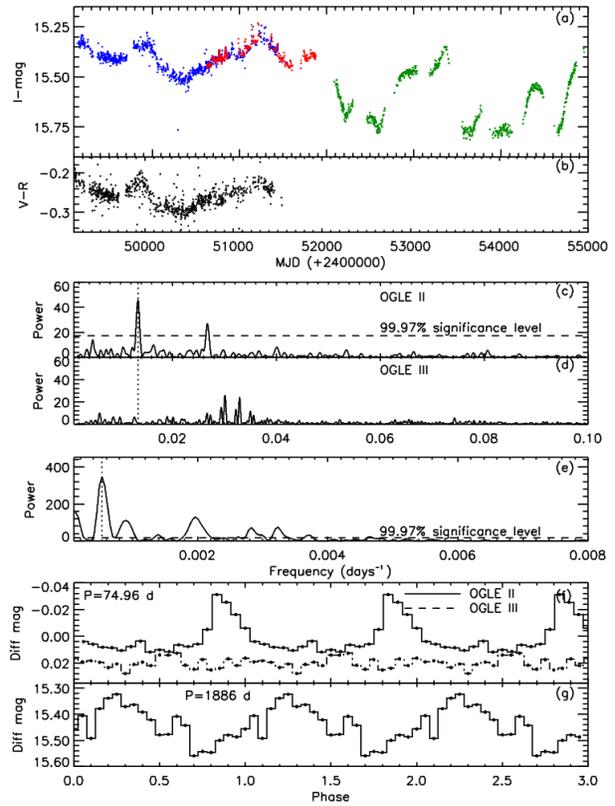}}
\caption{SXP455. (a): MACHO $\mathcal{R}$-band (blue), OGLE II (red), and OGLE III (green) light curves. (b): MACHO colour variation $\mathcal{V}$- $\mathcal{R}$. (c), (d), and (e): Power spectrum of the detrended OGLE II, OGLE III and combined light curves showing peaks at 74.96 d and 1886 d respectively. (f) and (g): Folded light curve on the presumed orbital and superorbital periods.}
\label{sxp455lc} 
\end{figure}

The light curve of SXP455 displays a substantial long-term variation with a period of 1886$\pm$145 d (see power spectra in Figure~\ref{sxp455lc}e). The light curve folded on this superorbital period is shown in Figure~\ref{sxp455lc}g.

The power spectrum of our detrended OGLE II light curve (Figure~\ref{sxp455lc}c) shows a peak at the 74.96$\pm$0.05 d orbital period. Remarkably, this periodicity appears absent in the OGLE III light curve (Figure~\ref{sxp455lc}d). The OGLE II and OGLE III light curves folded on this period are shown in Figure~\ref{sxp455lc}f and that of OGLE II exhibits a classic BeX outburst profile with a secondary peak.

\subsection{SXP504 (AX J0054.8-7244)}

The pulse period of SXP504 was independently discovered in Chandra archival data by \citet{edge04} with a period of $\sim$500s and in XMM-Newton data by \citet{haberl04} with a period of 499.2$\pm$0.7 s. The precise X-ray position is coincident with an emission-line star MA[93]809 which is the same star as 207.16254.16 in the MACHO catalogue. \citet{edge04} reported a period of 268.6$\pm$0.1 d in the MACHO and OGLE II data, which was presumed to be orbital.

\begin{figure}
\scalebox{0.46}{\includegraphics{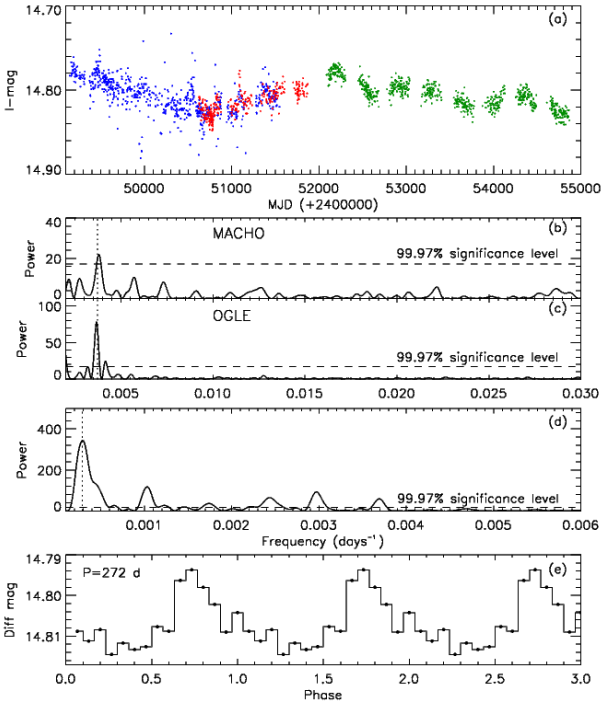}}
\caption{SXP504. (a): MACHO $\mathcal{R}$-band (blue), OGLE II (red), and OGLE III (green) light curves. (b) and (c): Power spectrum of the MACHO and OGLE III data showing a peak at 272 d. (d): Power spectrum of the combined light curve with a peak at 3448 d. (e): The folded light curve on the 272 d presumed orbital period.}
\label{sxp504lc} 
\end{figure}

The MACHO and OGLE light curves show clearly the regular optical outbursts at $\sim$ 270 d intervals. After detrending the MACHO and OGLE light curves, we computed their power spectra. A peak at 272$\pm$0.7 d was found in both MACHO, OGLE II and OGLE III data, which is consistent with the presumed orbital period (Figure~\ref{sxp504lc}b and \ref{sxp504lc}c). The folded OGLE III light curve on this period is shown in Figure~\ref{sxp504lc}e, which has the same profile as the MACHO folded light curve reported by \citet{edge04}.

The light curve exhibits a smooth long-term quasi-periodic modulation on a timescale of 3500 d. Accordingly, the L-S periodogram and PDM reveals very strong low-frequency power with a peak at 3448$\pm$119 d (Figure~\ref{sxp504lc}d).

\subsection{SXP565 (CXOU J005736.2-721934)}

\citet{macomb03} discovered the 564.8 s pulsations of this source with Chandra. The optical counterpart was identified as an emission-line star, [MA93] 1020 in the \citet{meys93} catalogue, and a period of 95.3 d was found by \citet{schmidtke04}. However, using RXTE/PCA observations, \citet{galache08} reported a period of 151.8 d.

\begin{figure}
\scalebox{0.46}{\includegraphics{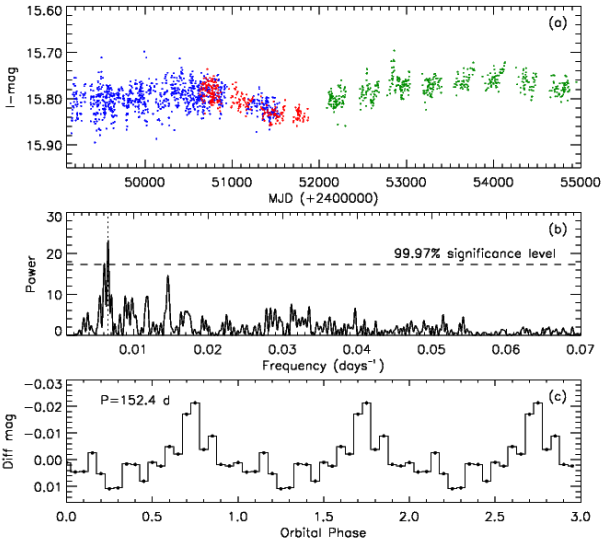}}
\caption{SXP565. (a): MACHO $\mathcal{R}$-band (blue), OGLE II (red) and OGLE III (green) light curves. (b): Power spectrum of the detrended light curves showing a significant peak at 152.4 d period. (c): The detrended light curve folded on the presumed 152.4 d orbital period.}
\label{sxp565lc} 
\end{figure}

To try to determine the real orbital period of the system, we have detrended the OGLE III data. Analysis of the detrended light curve reveals a peak at 152.4 $\pm$ 0.2 d (Figure~\ref{sxp565lc}b), which appears to be caused by several BeX-type outbursts in the light curve. The folded light curve on this period (Figure~\ref{sxp565lc}c) shows a profile similar to those seen in other BeX sources. This suggests that the 152.4 d period is the real orbital period of SXP565 which is in good agreement with the X-ray period of 151.8 d found by \citet{galache08}.

The combined long-term light curve (Figure~\ref{sxp565lc}a) shows a very clear quasi-periodic variation on a timescale of $\sim$3000 d.

\subsection{SXP645 (XMMU J005535.2-722906)}

This transient X-ray source was dectected with a pulse period of 645 s by \citet{haberl08a}. They identified the optical counterpart as a V$\sim$14.6 star which is object 207.16315.28 in the MACHO catalogue, and SMC-SC7 137527 in OGLE II.

\begin{figure}
\scalebox{1}{\includegraphics{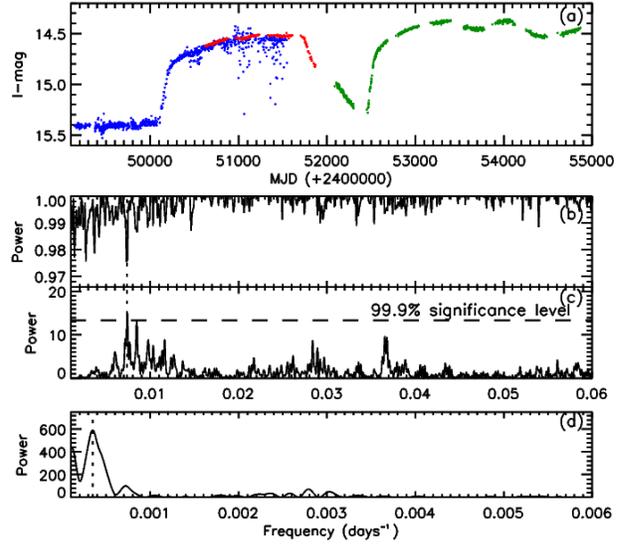}}
\caption{SXP645. (a): MACHO $\mathcal{R}$-band (blue), OGLE II (red) and OGLE III (green) light curves. (b), (c), and (d): Power spectrum (PDM and L-S)of the detrended and combined light curves showing peaks at 135.4 d and 2857 d.}
\label{sxp645lc} 
\end{figure}

The long-term light curves show very large amplitude ($\sim$ 1 mag) variations repeating after $\sim$ 2500 d (Figure~\ref{sxp645lc}a), which causes a strong low-frequency peak at 2857$\pm$81 d in the power spectrum (Figure~\ref{sxp645lc}d).

At higher frequency, we found a significant peak at 135.4$\pm$0.5 d in the PDM and L-S power spectrum (Figure~\ref{sxp645lc}b and ~\ref{sxp645lc}c), which could be the orbital period of SXP645.

\subsection{SXP701 (XMMU J005517.9-723853)}

\citet{haberl04} discovered 701.6$\pm$1.4 pulsations from this source during an XMM-Newton Target of Opportunity (ToO) observation of the SMC field around XTE J0055-727. This X-ray source is identified  with MACHO 207.16313.35 and OGLE II SMC-SC7 129062. \citet{schmidtke05b} found a weak 412 d modulation in the MACHO data.

\begin{figure}
\scalebox{0.46}{\includegraphics{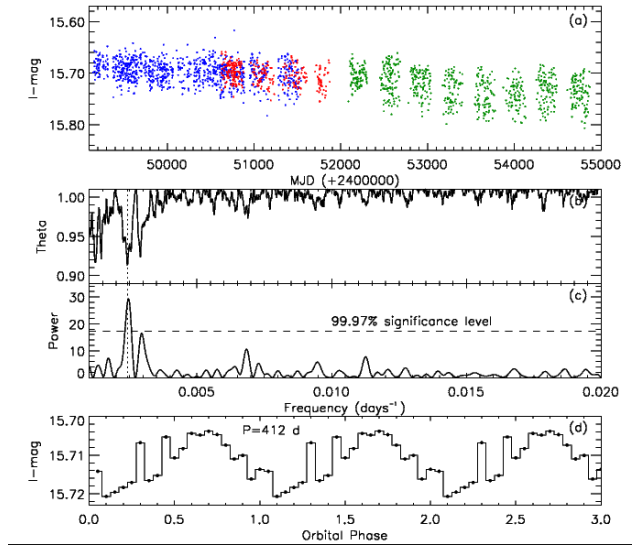}}
\caption{SXP701. (a): MACHO $\mathcal{R}$-band (blue), OGLE II (red) and OGLE III (green) light curves. (b) and (c): Power spectrum (PDM and L-S) of the detrended combined light curves showing a significant peak at 412 d period. (c): The detrended light curve folded on the 412 d period.}
\label{sxp701lc} 
\end{figure}

The power spectra (PDM and L-S) of our OGLE III light curve reveals a significant peak at 412$\pm$5 d, which is the presumed orbital period (Figure~\ref{sxp701lc}b, \ref{sxp701lc}c), very close to that found in MACHO data by \citet{schmidtke05b}. The folded light curve on this period is shown in (Figure~\ref{sxp701lc}d).

The long-term light curves do not show periodic variability other than the presumed orbital period, but a gradual fading over $\sim$ 5000 d (Figure~\ref{sxp701lc}a) is evident.

\subsection{SXP755 (RX J0049.7-7323)}

\citet{yokogawa00c} discovered 755.5$\pm$0.6 s pulsations in an ASCA observation of AX J0049.5-7323. The optical counterpart was identified as a V=15 star by \citet{edge03a}, and is designated [MA93] 315. The X-ray period of 396$\pm$5 d was found by \citet{laycock05} and is consistent with the optical period of $\sim$394 d reported by \citet{schmidtke04}.

\begin{figure}
\scalebox{0.46}{\includegraphics{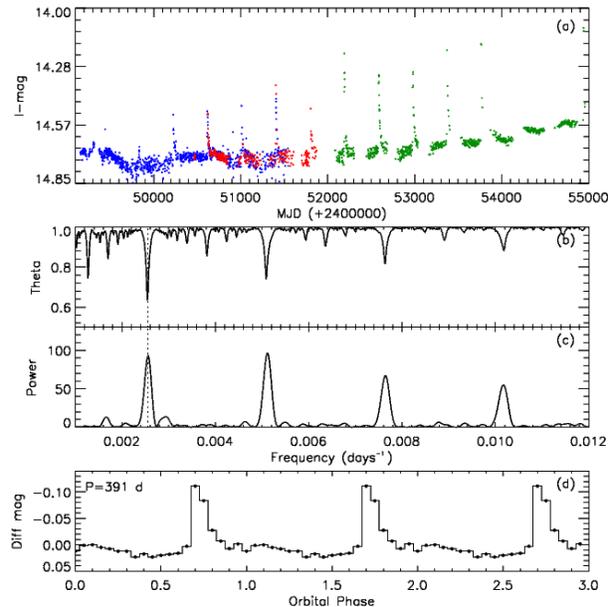}}
\caption{SXP755. (a): MACHO $\mathcal{R}$-band (blue), OGLE II (red) and OGLE III (green) light curves. (b) and (c): Power spectrum (PDM and LS) of the combined light curves showing a significant peak at 391 d and harmonics. (d): data folded on the 391 d period.}
\label{sxp755lc} 
\end{figure}

The light curve of SXP755 shows a series of outbursts which appear to strengthen as the source brightens (Figure~\ref{sxp755lc}a). The power spectrum (PDM and L-S) of the detrended combined light curve shows a peak at 391$\pm$2 d (Figure~\ref{sxp755lc}b, \ref{sxp755lc}c), very close to the presumed orbital period . The folded light curve on this period is shown in Figure~\ref{sxp755lc}d.

The combined light curves exhibit a gradual brightening over the $\sim$ 5000 d observation timescale (Figure~\ref{sxp755lc}a).

\section{Variation of outburst amplitude}\label{outbamp}

In addition to the regular orbital modulation, we have also seen variations in the amplitude of this modulation throughout associated long-term cycles of variability. Similar behaviour has already been reported in BeX sources such as A0588-66 \citep{mcgowan03} and SXP755 \citep{schmidtke04}. It has been suggested that the strength of the outbursts may depend on the density and extent of the equatorial disc at the time of periastron passage \citep[see][]{mcgowan03,schmidtke04}.

\begin{figure*}
\scalebox{0.86}{\includegraphics{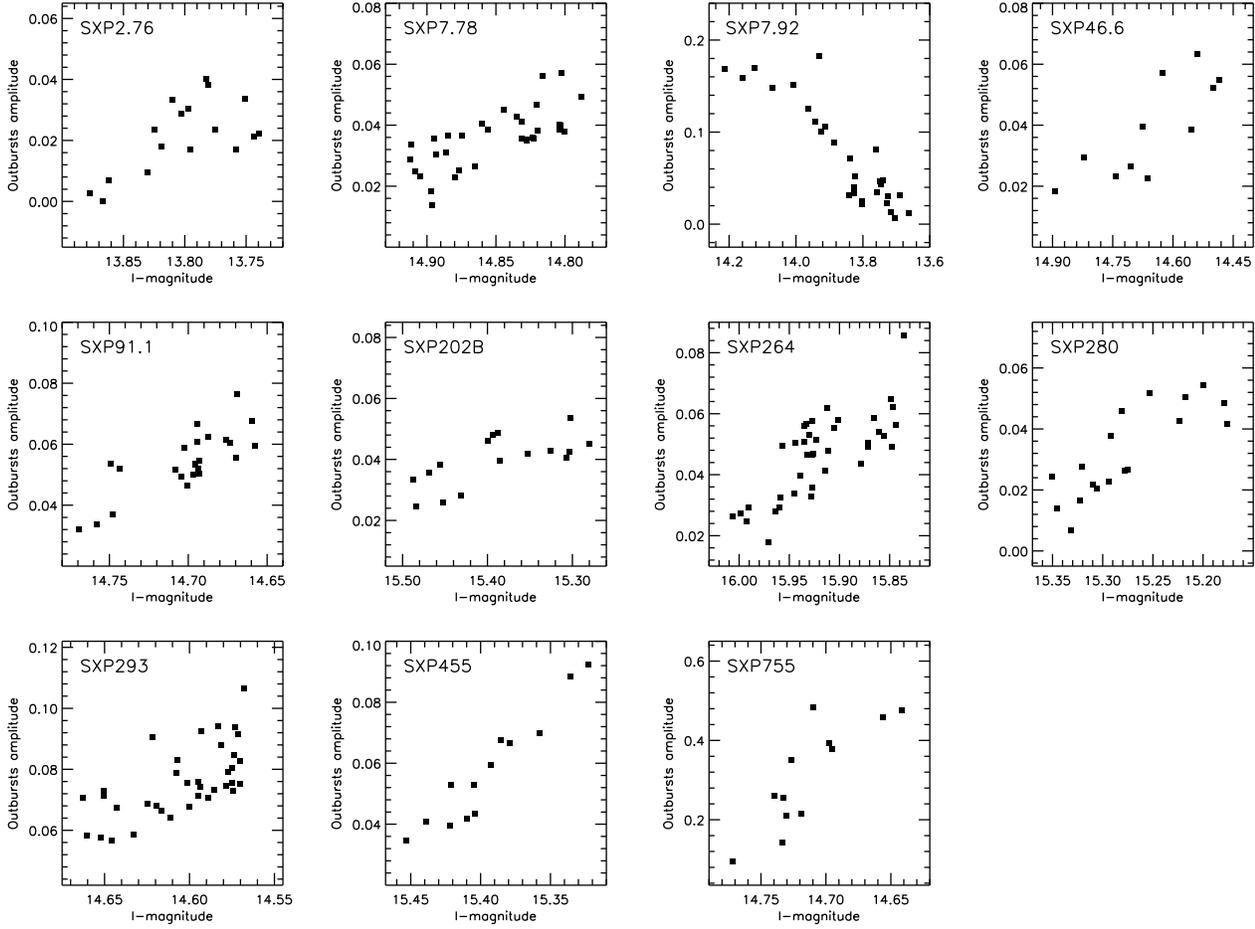}}
\caption{Variation of outburst amplitude of selected SMC BeX sources as a function of the brightness of the sources.}
\label{outburstvariation} 
\end{figure*}

For the sources considered here, we investigate this relationship through figure~\ref{outburstvariation} which plots the orbital outburst amplitude against the source brightness (presumed related to the size of the Be circumstellar disc). When the disc is larger (hence brighter), the neutron star's eccentric orbit leads to a more significant interaction (and hence outburst) during periastron passage. Conversely, when smaller, the disc interaction is weaker or absent completely.

Most of the SMC BeX sources exhibit an increase of outburst amplitude with brightness. The only exception is SXP7.92, which shows a decrease in strength of the outbursts when the source brightens. We suggest an explanation for this behaviour in section \ref{superorbital}.

\begin{figure*}
\scalebox{0.86}{\includegraphics{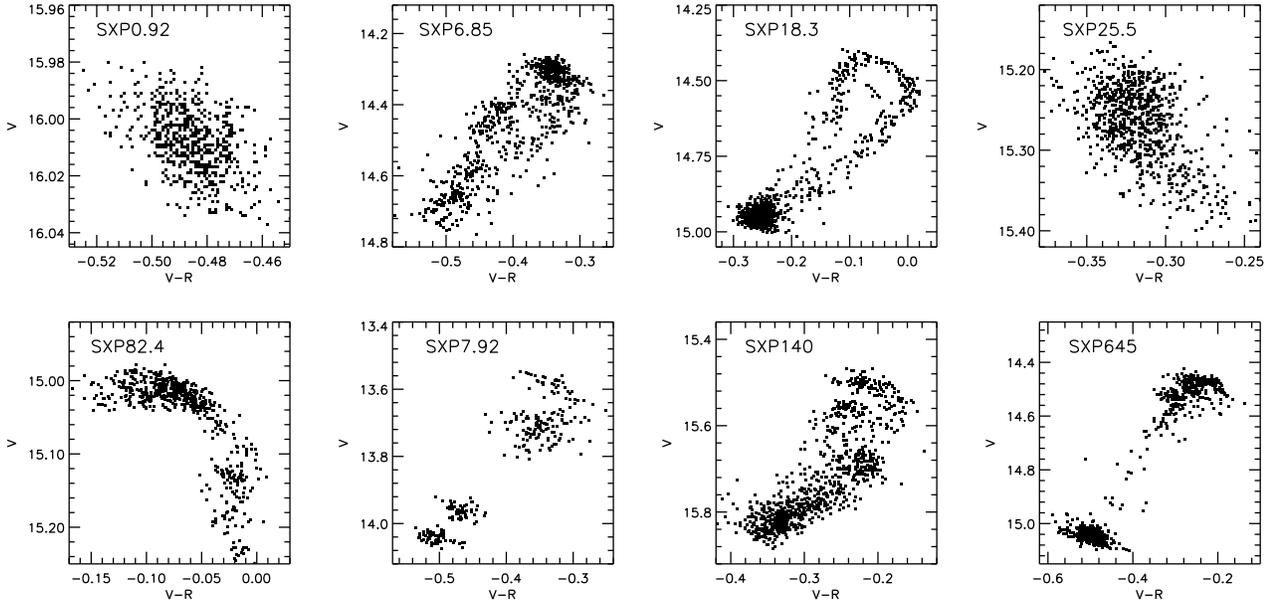}}
\caption{$\mathcal{V}$ vs. $\mathcal{V}$-$\mathcal{R}$ diagram of selected SMC BeX sources showing that they are redder when brighter; the 3 exceptions being SXP0.92, SXP25.5, and SXP82.4 which are redder when fainter.}
\label{colormag} 
\end{figure*}

\section{General discussion.}

In this paper we have exploited the $\sim$16yr timebase of the MACHO and OGLE archives in order to investigate the long-term variability properties of almost all Be/X-ray pulsars currently known in the SMC.  Virtually all of them show superorbital variations on timescales of hundreds to thousands of days.  While such variations had been known historically for some of the bright, galactic Be systems, this is the first systematic study on these timescales of the BeX SMC population, and it is clear that such variability is extremely common, much of it appearing quasi-periodic in nature.  

Optical light curves of these BeX systems show a variety of periodicities from as short as $\sim$hours (likely non-radial pulsation of the Be star) to tens of days (likely orbital modulation resulting from the neutron star's regular interaction with mass lost from the Be donor in its eccentric orbit) to many hundreds of days, which we presume to be a property of the circumstellar disc surrounding the Be donor.  This latter timescale is seen as both a low and high amplitude modulation, explanations for which range from an oscillation within the disc, to its formation and depletion.  We will discuss these properties in more detail.

The list of all superorbital and orbital periods found in this report as well as the previously reported X-ray and optical periods are combined in table~\ref{tab:listres}.

\subsection{Superorbital periods in BeX systems}\label{superorbital}

Long-term non-orbital variations have previously been seen in some BeX systems, such as the 421 d quasi-periodic modulation in A0538-66 \citep{alcock01}. This modulation was the driving force behind the current paper, as it suggested as an observable effect of variations in size of the circumstellar disc around the Be star. \citet{alcock01} reported the additional remarkable property that, in A0538-66, the strength of the 16.5 d orbital outbursts depend very strongly on the brightness of the source and hence are also a function of the 421 d cycle. They occur during optical minimum, but not when the star is at its maximum brightness. For this system, the decline in brightness has been suggested to be related to the formation of an extended, cool, equatorial disc about the Be star,  which will mask part of the Be star itself. A0538-66 becomes redder at minimum and bluer at optical maximum, which is what might be expected for a high inclination system. In contrast, we find here that only a few of the SMC BeX sources exhibit this type of behaviour, such as SXP25.5 (see figure~\ref{colormag}).

However, for almost all of the SMC sources examined here, the source gets {\it redder} when it brightens (figure~\ref{colormag}). For these systems, we interpret the red continuum as arising from the Be circumstellar disc seen at moderate or low inclination, i.e. the disc adds cooler light to that of the B star.  BeX systems can exhibit this type of behaviour if the inclination of the Be equatorial disc is $ \leq90^{o}-\alpha $, where $\alpha$ is the opening angle of the disc (and is likely very small, $\leq$ few degrees). Assuming that all the BeX binaries are oriented at random, the probability of a system having an inclination $i\leq90^{o}-\alpha$ is simply $1 - cos(90^{o}-\alpha)$. If we assume that the opening angle of the disc for a typical BeX source is $\alpha\sim10^{o}$, then the probability of a BeX having an inclination $<80^{o}$ is 0.83. i.e. we would therefore expect the BeX systems to exhibit this type of behaviour rather than that of A0538-66. In our sample of 31 sources with MACHO counterparts we would then expect only 4 or 5 sources to exhibit the ``A0538-type'' of behaviour, and this is consistent with the number (3 sources) that become redder as they get fainter (see figure~\ref{colormag}).

In addition, we have seen a correlation between the amplitude of outburst and brightness in some sources (such as SXP7.78, SXP293, SXP755, ...). For those systems, the amplitude of outburst varies as a function of the brightness of the source, the amplitude growing with the source brightness (see figure~\ref{outburstvariation}). This suggests that the size of the Be equatorial disc influences the scale of the interaction between the neutron star and the Be equatorial disk.

However, if the Be equatorial disc becomes sufficiently large, then it can influence the entire orbit of the neutron star, producing a type II outburst. In this case, the Be circumstellar disk almost reaches its maximum size because it is then truncated by the neutron star. At this point the neutron star acts to increase the density of the disc rather than extend its size, hence the optical modulation is greatly reduced. This would explain the strange behaviour of SXP7.92. In Figure~\ref{colormag}, the MACHO colour variation of SXP7.92 increases with the brightness of the source. However, its outburst amplitude becomes very weak as it brightens (Figure~\ref{outburstvariation}). For this system, the equatorial disc has become very large (it increases in brightness by $\sim$1 mag) and likely encompasses the entire orbit of the neutron star.

For highly variable sources (such as SXP6.85, SXP8.88, SXP15.3, SXP18.3,...), the MACHO colour variations follow the variation in brightness of the source. This confirms that the long-term variation in the light curve is related to the behaviour of the Be circumstellar disc.

\subsubsection{Contribution of the extended disc.}

Let us consider one of these highly variable SMC BeX sources. Assuming that SXP6.85 is viewed face-on ({\it i$\sim$0$^{\circ}$}) and it has no disc during optical minimum, then the 0.7 mag global change in the light curve represents the brightness contribution from the Be disc. This implies a disc luminosity of $\sim 3.3 \times 10^{35}~erg.s^{-1}$, which corresponds to a total increase in brightness by $\sim$95\% from the optical minimum ($L_{min}\sim 3.5 \times 10^{35}~erg.s^{-1}$). However, for a very high inclination system such as A0538-66 \citep[$i\geq 74^{\circ}.9\pm6^{\circ}.5$, ][]{mcgowan03}, the quiescent state (no disc, only the naked B star) corresponds to the maximum optical brightness, and then when the equatorial disc grows it masks part of the Be star and reduces its optical brightness. The change in brightness in the light curve of A0538-66 is about 0.5 mag. However, the size of the equatorial disc of this source cannot be large because of its short orbital period ($\mathrm {P_{orb}}$=16.65 d, \citealt{mcgowan03}), as the disc will be truncated at a smaller radius. The change in luminosity between no disc and maximum size of disc is about $2.7 \times 10^{35}~erg.s^{-1}$, this corresponds to a decrease in the total brightness of $\sim$37\% from its quiescent state.

\subsubsection{Interaction with the neutron star.}

It is interesting to note that, in spite of their long orbital periods, the presumed high orbital eccentricity leads to high neutron star velocities during periastron passage. Consider a typical BeX source consisting of a 1.4 $M_{\odot}$ compact object orbiting a 10 $M_{\odot}$ Be star with an eccentricity of {\it e = 0.7} and orbital periods of $P_{orb}$= 17 d and 300 d. At periastron (separation=  $1.3$ and $8.9 \times 10^{7}~km$), the neutron star would have a velocity of 443 and 170 $km.s^{-1}$ respectively. The projected equatorial rotational velocity of a Be star ($v~sin~i$) can be estimated from the broadening of its spectral lines, with a mean value of about $250~km.s^{-1}$ \citep{slettebak82}, which is comparable to the neutron star velocity at periastron. Therefore, at periastron, the primary star is rotating close to corotation with the orbiting neutron star.

\subsubsection{One-armed oscillation in the equatorial disc.}

Be stars are believed to exibit a cyclic variation in their emission line profiles (known as the `V/R' variations; \citealt{McLaughlin61}). It is widely accepted that these variations are caused by global one-armed oscillations in the equatorial discs of Be stars \citep{Kato83, Okazaki91}. The period of this long-term  V/R variation is typically in the range of 2-15 years for isolated Be stars \citep[eg.][]{Papaloizou92}. We note that these timescales are very similar to those seen in our light curves (see Table~\ref{tab:listres}). This suggests that the long-term variation in Be star emission line profiles (the `V/R' variation) may be related to the long-term variations we see in the optical light curves. Unfortunately, there are not yet any long-term spectrocopic studies of these SMC BeX sources capable of investigating the correlation between these two (photometric and spectrocopic) long-term variations.

The peaks and dips seen in SXP2.37 may possibly be explained as a direct effect of such a precessing elongated disk. For SXP2.37, the outbursts are seen during optical maximum, and then change into dips as the source fades. If we assume that SXP2.37 is a high inclination system, the source is brighter when the elongated disc extends to each side of the Be star, in this case the interaction between the neutron star and the disc can be seen as a maximum. On the other hand, when the elongated disc passes in front it will obscure the Be star, reducing its brightness. In this case, the neutron star passage will enlarge the disc which blocks some light from the hotter Be star, which we see as dips.

\subsection{Orbital and Super-orbital period correlation}

Apart from the very long-term ($\sim$ hundreds of days) variations that are of primary interest here, we have also seen optical orbital modulations in the light curves of these SMC BeX systems, visible as a series of regular outbursts. These outbursts are interpreted as the periastron passage of the neutron star where it interacts with the Be equatorial disc. At periastron, the circumstellar disc can be perturbed from its stable resonant state, this perturbation will slightly increase its surface area and, hence, its optical brightness \citep{okneg01}.

The optical light curves folded on the orbital period have an asymmetric profile, with a faster rise and slower decline. This behaviour has been seen in other BeX systems \citep{alcock01}. Furthermore, in some sources such as SXP46.6, SXP327, SXP348, etc., the optical outburst is not just a single maximum, but can show two peaks every binary cycle, which is seen in the light curves when folded on the orbital period. \citet{mcgowan07}, and \citet{coe09} suggested this behaviour as arising from the misalignment between the spin axis of the Be equatorial disc and the orbital plane of the binary system. Therefore, the neutron star interacts twice with the Be equatorial disc every binary cycle. We note that for SXP18.3, we find an optical periodicity (P= 28.5 d) that is distinct from its presumed orbital period (P= 17.92 d), which is clearly seen only in the MACHO, OGLE II, and first year of the OGLE III light curves. The mechanism that produces this 28.5 d optical modulation is not understood but it is clearly visible in the light curve. We note that it appears especially during the optical minimum.

The observed misalignment can be caused by the supernova kick received by the system when the neutron star was born. \citet{brandt95} suggested that an asymmetric supernova explosion can give very large kicks to the newly formed neutron stars which can either disrupt the system if the kick is too strong or lead to a large eccentricity and misalignment between the old and new orbits if the velocity of the kick is smaller. \citet{martin09} suggested that a velocity kick of 265 $km.s^{-1}$ is consistent with the observed misalignments in BeX systems, but too high for the observed eccentricities.

\begin{figure}
\scalebox{0.46}{\includegraphics{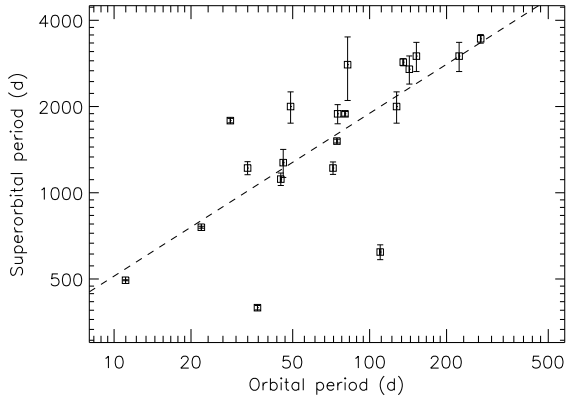}}
\caption{Plot of the BeX superorbital periods found in this work against orbital period. The dashed line represents the best linear fit.}
\label{PorbPsup} 
\end{figure}

In Figure~\ref{PorbPsup}, we have plotted the observed long-term superorbital period of the SMC BeX systems against their orbital periods. It appears that the two periods correlate. We have computed the linear Pearson correlation coefficient to be 0.73 with a p-value of 0.00039 (significant at the 99.9\% confidence level). Some authors have already suggested that these long-term variations might be related to the orbital period \citep{reig05,coe05}, based on the H$\alpha$ EW - P$_{orb}$ relationship of \citet{reig97}, which is in good agreement with the disc truncation model of \citet{okneg01}. If the Be equatorial disc is truncated by the neutron star orbit, a source with a shorter period (large number of periastron passages) would have a smaller equatorial disc than a source with a wider and more eccentric orbit.

\begin{table*}
\centering
\caption{Periodicities found in SMC BeX sources.}
\vspace{0.5cm}

\scriptsize{
\setlength{\columnsep}{-10pt}
\begin{tabular}{lrcrccr}\hline\hline
\\
\multicolumn{1}{c}{\(\bf {Short}\)}&
\multicolumn{2}{c}{\(\bf {P_{sup}^{\star}}\)}&
\multicolumn{2}{c}{\(\bf {P_{orb}}\)}&
\multicolumn{2}{c}{\(\bf {Previously~reported }\)}\\
\multicolumn{7}{c}{}\\

\multicolumn{1}{c}{\(\bf {ID} \)}&
\multicolumn{1}{c}{\(\bf {Period} \)}&
\multicolumn{1}{c}{\(\bf {Semi-amplitude} \)}&
\multicolumn{1}{c}{\(\bf {Period(error)} \)}&
\multicolumn{1}{c}{\(\bf {Semi-amplitude} \)}&
\multicolumn{1}{r}{\(\bf {P_{X-ray^{\dagger}}} \)}&
\multicolumn{1}{r}{\(\bf {P_{opt}[ref]} \)}\\

\multicolumn{1}{c}{}&
\multicolumn{1}{c}{[d]}&
\multicolumn{1}{c}{[mmag]}&
\multicolumn{1}{c}{[d]}&
\multicolumn{1}{c}{[mmag]}&
\multicolumn{1}{c}{[d]}&
\multicolumn{1}{c}{[d]}\\

\\
\hline
\\

SXP0.09	&	[247]		&	$<$5	&	...		&	$<$5	&    	...	&	...		\\
SXP0.92	&	2654$\pm$298	&	6	&	...		&	$<$5	&    	...	&	51[1]		\\
SXP2.37	&	...		&	$<$5	&	18.58(1)$^{\star \star}$	&	16	&	...	&	...		\\
SXP2.76	&	2800$\pm$700	&	73	&	82.37(7)	&	11	&	...	&	82.1[2]		\\
SXP3.34	&	[495]		&	$<$5	&	11.09(1)	&	14	&	...	&	...		\\
SXP6.85	&	621$\pm$4	&	201	&	110.0(2)	&	26	&	112	&	114[4]		\\
SXP7.78	&	1116$\pm$56	&	50	&	44.9(2)		&	26	&	44.9	&	44.8[5]		\\
SXP7.92	&	397$\pm$2	&	138	&	36.41(2)	&	15	&	...	&	36.8[14]	\\
SXP8.9	&	1786$\pm$32	&	485	&	28.51(1)	&	8	&	28.4	&	33.4[6]		\\
SXP9.13	&	1886$\pm$35	&	32	&	80.1(1)		&	8	&	77.2	&	40.1[8] 	\\
SXP15.3	&	1515$\pm$23	&	55	&	74.51(5)	&	12	&	28	&	75.1[8]		\\
SXP18.3	&	...		&	$<$5	&	17.95(1)	&	12	&	17.7	&	17.7[15]	\\
SXP22.1	&	...		&	$<$5	&	75.97(6)	&	9	&	...	&	...		\\
SXP25.5	&	...		&	$<$5	&	22.50(1)	&	16	&	...	&	...		\\
SXP31.0	&	...		&	$<$5	&	90.5(1)		&	14	&	...	&	90.4[2]		\\
SXP34.1	&	...		&	$<$5	&	[598]		&	$<$5	&	...	&	...		\\
SXP46.6	&	...		&	$<$5	&	136.4(2)	&	9	&	137	&	137[9]		\\
SXP59	&	...		&	$<$5	&	62.10(4)	&	8	&	122	&	60.2[10]	\\
SXP74.7 &	1220$\pm$64	&	21	&	33.37(1)	&	12	&	61.6	&	33.4[11]	\\
SXP82.4	&	...		&	$<$5	&	171(1)		&	10	&	362	&	...		\\
SXP91.1	&	...		&	$<$5	&	88.3(1)		&	24	&	117	&	88.2[7]		\\
SXP101	&	758$\pm$6	&	14	&	21.95(1)	&	12	&	25.2	&	21.9[12]	\\
SXP138	&	2700$\pm$304	&	172	&	[143.1]		&	$<$5	&	103	&	122[6]		\\
SXP140	&	492$\pm$2.4	&	110	&	...		&	$<$5	&	...	&	197[6]		\\
SXP172	&	...		&	$<$5	&	67.88(4)	&	22	&	70	&	69.9[6]		\\
SXP202A	&	1220$\pm$61	&	100	&	71.98(5)	&	18	&	91	&	...		\\
SXP202B	&	$\sim$3000    	&	108	&	224(1)		&	15	&	...	&	...		\\
SXP264	&	$\sim$2000	&	47	&	49.06(2)	&	18	&	...	&	49.1[10]	\\
SXP280	&	$\sim$2000	&	88	&	126.4(2)	&	15	&	64.8	&	127[2]		\\
SXP293	&	...		&	$<$5	&	59.77(3)	&	35	&	151	&	59.7[7]		\\
SXP304	&	...		&	$<$5	&	[344]		&	$<$5	&	...	&	520[6]		\\
SXP327	&	1274$\pm$143	&	41	&	45.9(2)		&	120	&	...	&	45.9[16]	\\
SXP348	&	...		&	$<$5	&	94.4(1)		&	9	&	...	&	93.9[6]		\\
SXP455	&	1886$\pm$145	&	110	&	74.96(5)	&	21	&	...	&	74.7[7]		\\
SXP504	&	3448$\pm$119	&	32	&	272(1)		&	10	&	265	&	273[10]		\\
SXP564	&	$\sim$3000	&	69	&	152.4(2)	&	15	&	151	&	95.3[7]		\\
SXP645	&	2857$\pm$81    	&	460	&	[135.3]		&	$<$5	&	...	&	...		\\
SXP701	&	...		&	$<$5	&	412(5)		&	8	&	...	&	412[10]		\\
SXP755	&	...		&	$<$5	&	391(2)		&	66	&	389	&	394[7]		\\

\\
\hline
\multicolumn{7}{p{14cm}}{[1]: \citet{kaspi93}; [2]: \citet{schmidtke06}; [3]: \citet{coe05}; [4]: \citet{mcgowan08}; [5]: \citet{cowley04}; [6]: \citet{schmidtkecow06}; [7]: \citet{schmidtke04}; [8]: \citet{edge05a}; [9]: \citet{mcgowan08}; [10]: \citet{schmidtke05b}; [11]: \citet{schmidtke07b}; [12]: \citet{mcgowan07}; [13]: \citet{edge05b}; [14]: \citet{coe09}; [15]: \citet{schurch09}; [16]: \citet{coe08}}\\
\\
\multicolumn{7}{p{14cm}}{Periods in square brackets are marginally significant (see section 2.3).}\\
\multicolumn{7}{p{14cm}}{$^{\star\star}$ 1 $\sigma$ uncertainty in paratheses (units of last digit). }\\
\multicolumn{7}{p{14cm}}{$^{\star}$Superorbital period.}\\
\multicolumn{7}{p{14cm}}{$^{\dagger}$X-ray orbital period from \citet{galache08}.}
\label{tab:listres}
\end{tabular}
}
\end{table*}

\section{Summary}

In summary, we have studied the long-term properties of SMC Be X-ray binaries. We have found 19 superorbital periods which we suggest are related to properties of the Be circumstellar disc (see table~\ref{tab:listres}). In addition, we have compiled and updated the orbital periods of these BeX sources using the complete MACHO, OGLE II, and OGLE III databases.

Furthermore, the amplitude of these outbursts vary significantly through the superorbital cycle. They are very strong either at optical maxima or at optical minima, depending on the inclination of the source. For high inclination systems, the Be circumstellar disc will mask the hotter and bluer star, and their outburst amplitudes will be strong at optical minima.

\section*{ACKNOwLEGMENTS}

We are grateful to Luis Balona, Malcom Coe, Brian Warner, Matthew Schurch, Lee Townsend, Brian Van Soelen and Patricia Whitelock for useful discussions. We also thank the anonymous referee for useful suggestions, particularly in clarifying the significance or otherwise of the low amplitude modulation. AFR is supported financially by the South African SKA project.

This paper utilizes public domain data obtained by the MACHO Project, jointly funded by the US Department of Energy through the University of California, Lawrence Livermore National Laboratory under contract No. W-7405-Eng-48, by the National Science Foundation through the Center for Particle Astrophysics of the University of California under cooperative agreement AST-8809616, and by the Mount Stromlo and Siding Spring Observatory, part of the Australian National University. The OGLE project is partially supported by the Polish MNiSW grant N20303032/4275.

\label{lastpage}

\end{document}